\documentclass[pra,twocolumn,showpacs,superscriptaddress,floatfix,nofootinbib]{revtex4-1}
\usepackage{epsfig}
\usepackage{latexsym}
\usepackage{amsmath}
\usepackage{graphics}
\usepackage{float}
\usepackage{bbold}

\usepackage{color}
\usepackage{pstricks}
\definecolor{MyRed}{RGB}{153,0,0}

\newcommand{\br}{\mathbf{r}}

\newcommand{\bq}{\mathbf{q}}
\newcommand{\brp}{\mathbf{r}_\parallel}

\newcommand{\bqp}{\mathbf{q}_\parallel}
\newcommand{\TE}{{\rm TE}}
\newcommand{\TM}{{\rm TM}}
\newcommand{\T}{\textsf{T}}
\newcommand{\w}{\omega}
\newcommand{\wP}{\widetilde{\omega}_{\rm P\parallel}}
\newcommand{\rd}{\hspace{-1 mm}{ d}}
\begin{document}

\title{Quantum electrodynamics near anisotropic polarizable materials: \\Casimir-Polder shifts near multi-layers of graphene}
\author{Claudia Eberlein}
%\affiliation{Department of Physics \& Astronomy,
%    University of Sussex,
%    Falmer, Brighton BN1 9QH, England}
\author{Robert Zietal}
\affiliation{Department of Physics \& Astronomy,
    University of Sussex,
     Falmer, Brighton BN1 9QH, England}
\date{\today}
\begin{abstract}
In a recent paper we have formulated a theory of non-relativistic quantum electrodynamics in the presence of an inhomogeneous Huttner-Barnett dielectric. Here we generalize the formalism to anisotropic materials and show how it may be modified to include conducting surfaces. We start with the derivation of the photon propagator for a slab of material and use it to work out the energy-level shift near a medium whose conductivity in the direction parallel to the surface far exceeds that in the direction perpendicular to the surface. We investigate the influence of the anisotropy of the material's electromagnetic response on the Casimir-Polder shifts, both analytically and numerically, and show that it may have a significant impact on the atom-surface interaction, especially in the non-retarded regime, i.e.~for small atom-surface separations. Our results for the energy shift may be used to estimate the Casimir-Polder force acting on quantum objects close to multilayers of graphene or graphite. They are particularly important for the case of trapped cold molecules whose dispersive interactions with surfaces often fall within the non-retarded regime where the anisotropy of the material strongly influences the Casimir-Polder force. We also give a formula for the change in the spontaneous decay rate of an excited atom or molecule near an anisotropically conducting surface.
\end{abstract}


\maketitle
\section{\label{sec:level1}Introduction}
%\section{\label{sec:level2}Electromagnetic field near a polarizable body}
In a recent paper \cite{Dispersive1} we have described a microscopic model for the interaction of the quantized electromagnetic field with an absorptive dielectric body. The basic idea behind the model, first introduced in \cite{Huttner1}, is that a linearly responding dielectric material can be perceived as a continuum of damped quantum harmonic oscillators, making up a Huttner-Barnett dielectric. Then the classical response of the material to an externally applied electric field is described by the Drude-Lorentz permittivity,
\begin{equation}
\epsilon\left(\br,\w\right)=1+g(\br)\frac{\w_{\rm P}^2}{\w^2_{\rm T}-\w^2-2i\gamma\w}\label{eqn:Epsilon}.
\end{equation}
In Ref.~\cite{Dispersive1} we have demonstrated that the Hamiltonian density of such a system can be written as
\begin{eqnarray}
\mathcal{H}_{\rm EM}\;&=&\;\frac{1}{2\epsilon_0}\mathbf{D}^2(\br)+\frac{1}{2\mu_0}\mathbf{B}^2(\br),\label{eqn:EMHam}\\
\mathcal{H}_{\rm P}\;&=&\;\frac{\mathbf{P}^2(\br)}{2\mathcal{M}}+\frac{1}{2}\mathcal{M}\w_{\rm T}^2\mathbf{X}^2(\br),\label{eqn:PolarizationHam}\\
\mathcal{H}_{\rm R}\;&=&\;\int_0^\infty\rd \nu\left[\frac{\mathbf{Z}^2_\nu(\br)}{2\rho_\nu}+\frac{1}{2}\rho_\nu\nu^2\mathbf{Y}_\nu^2(\br)\right],\label{eqn:BathHam}\\
\mathcal{H}_{\rm P-R}\;&=&\;-\int_0^\infty\rd \nu\rho_\nu \nu^2\mathbf{X}(\br)\cdot\mathbf{Y}_\nu(\br),\label{eqn:BathResCoupling}\\
\mathcal{H}_{\rm P-EM}\;&=&\;-\frac{g(\br)}{\epsilon_0}\mathbf{D}(\br)\cdot\mathbf{X}(\br),
\label{eqn:PolEMCoupling}\\
\mathcal{H}_{\rm S}\;&=\;&\frac{1}{2}\int_0^\infty\rd \nu\rho_\nu\nu^2\mathbf{X}^2(\br)+\frac{1}{2}\frac{g^2(\br)}{\epsilon_0}\mathbf{X}^2(\br),\label{eqn:FrequencyShifts}
\end{eqnarray}
where $\mathcal{H}_{\rm EM}$ describes the dynamics of the electromagnetic field, $\mathcal{H}_{\rm P}$ that of the polarization field, $\mathcal{H}_{\rm R}$ that of the reservoir responsible for the absorption, $\mathcal{H}_{\rm P-R}$ couples the polarization field and the reservoir, $\mathcal{H}_{\rm P-EM}$ couples the polarization and the electromagnetic fields, and $\mathcal{H}_{\rm S}$ arises from the frequency-shifts due to these two couplings. 
In order for Eqs.~(\ref{eqn:EMHam})--(\ref{eqn:FrequencyShifts}) to be meaningful the following equal-time commutation relations are required:
\begin{eqnarray}
 \left[D_i(\br),B_j(\br')\right]\;\;&=\;\;&i\hbar\epsilon^{ijm}\nabla'_m\delta^{(3)}(\br-\br'),\label{eqn:EMCommutator}\\
 \left[X_i(\br),P_j(\br')\right]\;\;&=\;\;&i\hbar\delta_{ij}\delta^{(3)}(\br-\br'),\label{eqn:PolComm}\\
\left[Y_{i,\nu}(\br),Z_{j,\nu'}(\br')\right]\;\;&=\;\;&i\hbar\delta_{ij}\delta^{(3)}(\br-\br')\delta(\nu-\nu').\label{eqn:BathCommutator}
\end{eqnarray}
The field ${\mathbf{D}(\br)\equiv\epsilon_0\mathbf{E}(\br)+g(\br)\mathbf{X}(\br)}$ is the divergence-free displacement field. Its appearance is a consequence of the multipolar coupling between the electromagnetic field and the polarization field $\mathbf{X}(\br)$ \cite{Yeung}.
The polarization field is in turn coupled to a continuum of bath oscillators $\mathbf{Y}_\nu(\br)$ with variable inertia $\rho_\nu$ leading to the absorption of radiation in the model. The position-dependent coupling function $g(\br)$ characterizes the volume of space where the matter-radiation interaction is switched on. It is equal to unity in the volume of space occupied by the dielectric and zero otherwise. Therefore it determines the shape of the polarizable body in question, cf. Eq. (\ref{eqn:Epsilon}). For a more detailed description of the model we refer the reader to Ref.~\cite{Dispersive1}.

The Hamiltonian
\begin{equation}
H=\int\rd^3\br\left(\mathcal{H}_{\rm EM}+\mathcal{H}_{\rm P}+\mathcal{H}_{\rm R}+\mathcal{H}_{\rm P-R}+\mathcal{H}_{\rm P-EM}+\mathcal{H}_{\rm S}\right)\nonumber
\end{equation}
can be in principle diagonalized, see e.g. \cite{SuttorpWubs}, and the exact electromagnetic field operators can be written down for some simple geometries of the dielectric body. However, as we have shown in \cite{Dispersive1}, it is also possible to sidestep the somewhat complicated diagonalization procedure and instead use the well-known methods of quantum field theory to directly obtain the exact Feynman propagator for the displacement field. The set of Eqs.~(\ref{eqn:EMHam})--(\ref{eqn:FrequencyShifts}) define an interacting quantum field theory and a rather simple one where all the couplings are bilinear. The chief quantity of interest is the propagator of the displacement field, which couples to any additional electric dipole moments in the system and as such can be understood as mediating interactions between atoms and molecules through their (fluctuating or permanent) electric dipole moments. With knowledge of this propagator one can calculate various experimentally measurable quantities, e.g. the Casimir-Polder shift in a neutral atom or molecule, in much the same way as this is done in free-space quantum electrodynamics. The Feynman propagator is an auxiliary mathematical construct which happens to appear in the diagrammatic formulation of perturbative quantum field theory and in general it does not necessarily have a clear physical interpretation \cite{Dirac}. However, in our model we may still associate the propagation of photons with the displacement field propagator defined as
\begin{equation}
D_{ij}(\br,\br;t,t')=-\frac{i}{\hbar}\left\langle 0 \left| {\T}\left[D_i(\br,t)D_j(\br',t')\right] \right| 0 \right\rangle\; ,
\label{eqn:DProp}
\end{equation}
and therefore we shall call $D_{ij}(\br,\br;t,t')$ the photon propagator. The state $|0\rangle$ in Eq. (\ref{eqn:DProp}) denotes the exact ground state of the \emph{interacting} system, and $D_i(\br,t)$ is the displacement field operator in the Heisenberg picture whose dynamics is governed by the full complement of Hamiltonian densities (\ref{eqn:EMHam})--(\ref{eqn:FrequencyShifts}), that is to say, \emph{including all the couplings}. Using basic diagrammatic techniques we have shown in Ref.~\cite{Dispersive1} that the Dyson equation for the photon propagator can be written as
\begin{eqnarray}
&\;&D_{il}(\br,\br';\w)= D^{(0)}_{il}(\br-\br';\w)\nonumber\\
&+& \dfrac{K(\w)}{\epsilon_0^2}\int \rd^3 \br_1 g(\br_1)D^{(0)}_{ij}(\br-\br_1;\w)D_{jl}(\br_1,\br';\w)\;\;\;
\label{eqn:IntegralIsotropic}
\end{eqnarray}
where $D^{(0)}_{il}(\br-\br';\w)$ is the free-space photon propagator and $K(\w)$ is a frequency-dependent function related to the permittivity of the medium (see Sec.~IV of Ref.~\cite{Dispersive1} for details). The benefit of working with propagators is that with Feynman diagrams one has at one's disposal a very efficient bookkeeping device for keeping track of higher-order contributions to perturbation series, which otherwise is a non-trivial task. As an example of what quantum field theory can do for atomic physics consider the van der Waals interaction of two identical atoms with one of them excited. If the interaction energy is calculated by using fourth-order time-independent perturbation theory one needs to consider twelve graphs and the same number of contributing terms \cite{Power}, but in quantum field theory the calculation of the same quantity requires only two Feynman diagrams \cite{Janowicz}. 
%%%%%%%%%%%%%%%%%%%%%%%%%%%%%%%%%%%%%
\begin{figure}[ht]
\includegraphics[width=5 cm, height=6 cm]{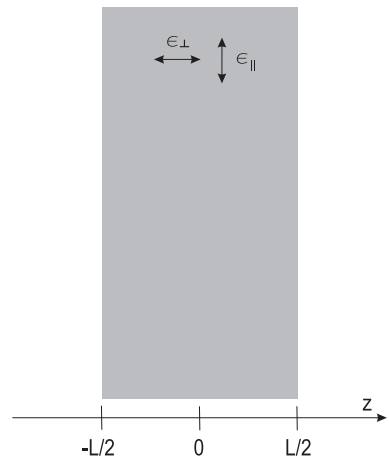}% Here is how to import EPS art
\caption{\label{fig:Slab} (Color online) The Feynman propagator for the displacement field, given in Eq. (\ref{eqn:SlabTotal}), describes the propagation of photons near a slab of absorptive material with anisotropic and frequency-dependent permittivity $\epsilon_\sigma(\w)$. It is calculated exactly starting from the quantum model for the light-matter interactions described in Secs. \ref{sec:level1} \& \ref{sec:level2}.}
\end{figure}
%%%%%%%%%%%%%%%%%%%%%%%%%%%%%%%%%%%%%%%%

It is well known that there are materials whose response to electromagnetic fields is anisotropic i.e.~which depends on the direction of the applied field. Imagine e.g.~a slab of the material, cf.~Fig.~\ref{fig:Slab}, which in the direction parallel to its surfaces behaves as a good conductor but in the direction perpendicular to its surfaces is a dielectric (or weak conductor) \cite{fn1}. An example of such a material would be a single layer or multi-layer films of graphene, which have recently started to play a significant role in nanotechnology, nanoelectronics and even experimental cold-atom physics. Thus, it seems to be of great importance to have appropriate tools at hand for the calculation of radiative corrections to properties of quantum systems, like atoms and molecules, near such anisotropic interfaces. The main goal of this paper is to provide those tools. In what follows we are going to show how the Huttner-Barnett model described above can be generalized to anisotropic dielectric media. As an example we will calculate the Casimir-Polder force acting on an atom close to an anisotropic dielectric slab. This paper will heavily rely on previous work reported in Ref.~\cite{Dispersive1}, from where we will be quote results instead of re-deriving them.  So, this paper should be read in conjunction with Ref.~\cite{Dispersive1}, in particular with Sections III and IV thereof.

\section{\label{sec:level2}Photon propagator near anisotropic media}
For an anisotropic dielectric slab, as depicted in Fig. \ref{fig:Slab}, the permittivity needs to be generalized from a scalar to a tensor
\begin{equation}
\boldsymbol{\epsilon}(z,\w)=\left(\begin{array}{ccc}
\epsilon_\parallel(z,\w) & 0 & 0 \\
0 & \epsilon_\parallel(z,\w) & 0 \\
0 & 0 & \epsilon_\perp(z,\w)
\end{array}\right),\label{eqn:DielMatrix}
\end{equation} 
where $\epsilon_\perp$ and $\epsilon_\parallel$ are still of the Drude-Lorentz form of Eq.~(\ref{eqn:Epsilon}) and, for a slab, depend only on $z$ coordinate. In order to achieve a dielectric response of the form (\ref{eqn:DielMatrix}) in our model we introduce an anisotropic polarization field and replace  Eq. (\ref{eqn:PolarizationHam}) with
\begin{eqnarray}
\mathcal{H}_{\rm P}&=&\frac{\mathbf{P}_\parallel^2(\br)}{2\mathcal{M_\parallel}}+\frac{1}{2}\mathcal{M_\parallel}\w_{\rm T\parallel}^2\mathbf{X}_\parallel^2(\br)\nonumber\\
&\;&+\frac{P_\perp^2(\br)}{2\mathcal{M_\perp}}+\frac{1}{2}\mathcal{M_\perp}\w_{\rm T\perp}^2X_\perp^2(\br)\; . \label{eqn:AnisotropicPolarizationHam}
\end{eqnarray}
To account for the slab geometry we fix the coupling function in Eqs. (\ref{eqn:PolEMCoupling}) and (\ref{eqn:FrequencyShifts}) to
\begin{equation}
g(\br)=1-\theta\left(-z-\frac{L}{2}\right)-\theta\left(z-\frac{L}{2}\right)\label{eqn:coupling},
\end{equation}
as appropriate for a dielectric slab of width $L$ centred on the $z=0$ plane, as shown in Fig. \ref{fig:Slab}. Here $\theta(z)$ is the Heaviside step function.  

In order to derive a generalization of the integral equation (\ref{eqn:IntegralIsotropic}) that reflects the anisotropy of the dielectric, one needs to successively integrate out the matter degrees of freedom from the Hamiltonian density. This can be done along exactly the same lines as described in Secs. III \& IV of Ref.~ \cite{Dispersive1}. The only difference is a change in the polarization field propagator on account of the Hamiltonian density having changed from Eq.~(\ref{eqn:PolarizationHam}) to Eq.~(\ref{eqn:AnisotropicPolarizationHam}). 
We find that the dressed polarization field propagator is still diagonal, but its diagonal elements are no longer equal as they were for an isotropic medium. Now we have
\begin{equation}
\boldsymbol{K}(\br-\br';\w)=\left(\begin{array}{ccc}
K_\parallel(\w) & 0 & 0 \\
0 & K_\parallel(\w) & 0 \\
0 & 0 & K_\perp(\w)
\end{array}\right)\delta^{(3)}(\br-\br')
\end{equation}
with
\begin{equation}
\left(1+\frac{K_\sigma(\w)}{\epsilon_0}\right)^{-1}=\xi_\sigma\left(\w\right)=1+\frac{\w_{\rm P\sigma}^2}{\w^2_{\rm T\sigma}-\w^2-2i\gamma_\sigma\sqrt{\w^2}}\label{eqn:Coincide}
\end{equation}
where $\sigma=\{\parallel, \perp\}$. Following the same steps as described in Secs. III \& IV of Ref.~\cite{Dispersive1}, we derive the generalized integral equation for the photon propagator,
\begin{eqnarray}
&\;&D_{il}(\br,\br';\w) = D^{(0)}_{il}(\br-\br';\w)\nonumber\\
&+& \frac{1}{\epsilon_0^2}\int \rd^3 \br_1 g(\br_1)D^{(0)}_{ij}(\br-\br_1;\w)K_{jk}(\w)D_{kl}(\br_1,\br';\w)\hspace{.7 cm}
\label{eqn:IntegralAnisotropic}
\end{eqnarray}
with the free-space photon propagator given by
\begin{equation}
D^{(0)}_{ij}(\br-\br';\w)=\frac{\epsilon_0}{(2\pi)^3}\int \rd ^3\bq\frac{\delta_{ij}\bq^2-q_iq_j}{\w^2-\bq^2+i\eta}\;e^{i\bq\cdot(\br-\br')}.\label{eqn:FreePropagator}
\end{equation}
The solution of Eq. (\ref{eqn:IntegralAnisotropic}), with $g(\br)$ as in Eq. (\ref{eqn:coupling}), is going to yield the photon propagator for the displacement field near an anisotropic dielectric slab, and with some skill it can be obtained by iteration. We are going to rely on a similar trick to the one used in Ref.~\cite{Dispersive1} for solving the equivalent integral equation for the geometry of a dielectric half-space. Equation (\ref{eqn:IntegralAnisotropic}), as it stands, cannot be iterated directly but can be reformulated in such a way that iteration is possible. For the slab geometry the matter-radiation interaction Hamiltonian reads
\begin{equation}
\mathcal{H}_{\rm P-EM}=-\frac{1}{\epsilon_0}\left[1-\theta\left(-z-\frac{L}{2}\right)-\theta\left(z-\frac{L}{2}\right)\right]\mathbf{X}(\br)\cdot\mathbf{D}(\br).\label{eqn:ExplicitCoupling}
\end{equation} 
There are several ways of splitting this Hamiltonian into an unperturbed part, for which the photon propagator is known, and one to treat as a perturbation, and different choices lead to different integral equations but for \emph{the same} photon propagator. Equation (\ref{eqn:IntegralAnisotropic}) has followed from taking the whole of Eq. (\ref{eqn:ExplicitCoupling}) as the perturbation and the unperturbed part of the Hamiltonian as that of the free electromagnetic field. However, we are free to split up the Hamiltonian differently, provided we know the photon propagator for the unperturbed part of the Hamiltonian. Let us do this in two different ways.
First we split the interaction Hamiltonian so as to include the polarization field for a dielectric half-space to the right of $-L/2$ in the unperturbed Hamiltonian,
\begin{eqnarray}
\mathcal{H}^{(0)}&=&\mathcal{H}_{\rm EM}-\frac{1}{\epsilon_0}\left[1-\theta\left(-z-\frac{L}{2}\right)\right]\mathbf{X}(\br)\cdot\mathbf{D}(\br),\nonumber\\
\mathcal{H}_{\rm P-EM}&=&\frac{1}{\epsilon_0}\theta\left(z-\frac{L}{2}\right)\mathbf{X}(\br)\cdot\mathbf{D}(\br).\label{eqn:ExplicitCoupling1}
\end{eqnarray} 
With this choice of splitting, the integral equation for the photon propagator reads
\begin{eqnarray}
&\;&D_{il}(\br,\br';\w) = D^{(+)}_{il}(\br,\br';\w)\nonumber\\
&-& \frac{1}{\epsilon_0^2}\int \rd^3 \br_1 \theta\left(z_1-\frac{L}{2}\right)D^{(+)}_{ij}(\br,\br_1;\w)K_{jk}(\w)D_{kl}(\br_1,\br';\w)
\nonumber\\
\label{eqn:IntegralAnisotropic1}
\end{eqnarray}
where $D^{(+)}_{il}(\br,\br';\w)$ is the photon propagator for the case of an anisotropic dielectric half-space occupying the region $z>-L/2$.
Next we reverse this choice and split the interaction Hamiltonian so as to include the polarization field for a dielectric half-space to the left of $L/2$ in the unperturbed Hamiltonian,
\begin{eqnarray}
\mathcal{H}^{(0)}&=&\mathcal{H}_{\rm EM}-\frac{1}{\epsilon_0}\left[1-\theta\left(z-\frac{L}{2}\right)\right]\mathbf{X}(\br)\cdot\mathbf{D}(\br),\nonumber\\
\mathcal{H}_{\rm P-EM}&=&\frac{1}{\epsilon_0}\theta\left(-z-\frac{L}{2}\right)\mathbf{X}(\br)\cdot\mathbf{D}(\br).\label{eqn:ExplicitCoupling2}
\end{eqnarray} 
With this choice of splitting, the integral equation for the photon propagator reads
\begin{eqnarray}
&\;&D_{il}(\br,\br';\w) = D^{(-)}_{il}(\br,\br';\w)\nonumber\\
&-& \frac{1}{\epsilon_0^2}\int \rd^3 \br_1 \theta\left(-z_1-\frac{L}{2}\right)D^{(-)}_{ij}(\br,\br_1;\w)K_{jk}(\w)D_{kl}(\br_1,\br';\w)
\nonumber\\
\label{eqn:IntegralAnisotropic2}
\end{eqnarray}
where $D^{(-)}_{il}(\br,\br';\w)$ is the photon propagator for the case of an anisotropic dielectric half-space occupying the region $z<L/2$. 

To keep the notation concise, we will from here on work with quantities that have been Fourier transformed in the direction parallel to the surface, e.g. with
\begin{equation}
D_{ij}(z,z';\bqp,\w)=\int\rd^2\mathbf{R}_\parallel e^{-i\bqp\cdot\mathbf{R}_\parallel}D_{ij}(\mathbf{R}_\parallel,z,z';\w).\label{eqn:FourierTransform}
\end{equation}
where $\mathbf{R}_\parallel=\brp-\brp'$. For notational clarity we will also suppress the dependence of the propagators on the frequency $\w$ and on the component $\bqp$ of the wave vector parallel to the interface.

We proceed by substituting Eq.~(\ref{eqn:IntegralAnisotropic2}) into the right-hand side of Eq.~(\ref{eqn:IntegralAnisotropic1}), which gives
\begin{eqnarray}
D_{ij}(z,z')&=&D_{ij}^{(+)}(z,z')\nonumber\\
&-&\frac{1}{\epsilon_0^2}\int_{L/2}^\infty\rd z_1D_{ik}^{(+)}(z,z_1)K_{kl}(\w)D_{lj}^{(-)}(z_1,z')\nonumber\\
&+&\frac{1}{\epsilon^4_0} \int_{L/2}^\infty \rd z_1 \int_{-\infty}^{-L/2}\rd z_2 D_{ik}^{(+)}(z,z_1)K_{kl}(\w)\nonumber\\
&\;&\times D_{lm}^{(-)}(z_1,z_2)K_{mn}(\w)D_{nj}(z_2,z').\label{eqn:IntegralFinalAnisotropic}
\end{eqnarray}
Integral equation (\ref{eqn:IntegralFinalAnisotropic}) is equivalent to Eqs.~(\ref{eqn:IntegralAnisotropic}), (\ref{eqn:IntegralAnisotropic1}), and (\ref{eqn:IntegralAnisotropic2}) in the sense that their solution is the same photon propagator for the anisotropic dielectric slab. The crucial difference is that the integral equation (\ref{eqn:IntegralFinalAnisotropic}) can be solved by iteration, whereas the others could not. 
To show that this is so, we need to know the photon propagator near an anisotropic dielectric half-space. The latter can be calculated by building on the expertise gained from the  calculation of the photon propagator near an isotropic dielectric half-space and reported in Ref.~\cite{Dispersive1}.

\subsection{Photon propagator near an anisotropic dielectric half-space}
Let us consider a half-space that occupies the region $z<0$. 
We seek the solution of Eq.~(\ref{eqn:IntegralAnisotropic}) with $g(\br)=\theta(-z)$, which can be found by iteration.  We only outline the calculation in this section, as we are closely following the steps of Sec. IV B of Ref.~\cite{Dispersive1},  and we shall concentrate on the modifications needed to account for the anisotropy of the material. The calculations are very similar, although a little bit more tedious. For a half-space filling the region $z<0$, the photon propagator satisfies the following set of integral equations \cite{Dispersive1},
\begin{eqnarray}
D_{il}(z,z')=D^{(0)}_{il}(z-z')\hspace{3.8 cm}
\nonumber\\
+\frac{1}{\epsilon_0^2}\int_{-\infty}^0 \rd z_1 D^{(0)}_{ij}(z-z_1)K_{jk}(\w)D_{kl}(z_1,z').\;\;\;\;\;\;
\label{eqn:Halfspace1}\\
D_{il}(z,z')=D^{(\epsilon)}_{il}(z-z')\hspace{3.7 cm}\nonumber\\
-\frac{1}{\epsilon_0^2}\int_{0}^\infty \rd z_1 D^{(\epsilon)}_{ij}(z-z_1)K_{jk}(\w)D_{kl}(z_1,z'),\;\;\;\;\;
\label{eqn:Halfspace2}
\end{eqnarray}
were $D^{(0)}_{il}(z-z')$ and $D^{(\epsilon)}_{il}(z-z')$ are the propagators in free space and in a bulk (now anisotropic) medium, respectively.  Equations (\ref{eqn:Halfspace1}) and (\ref{eqn:Halfspace2}) combine to yield
\begin{eqnarray}
D_{in}(z,z')=D_{in}^{(\epsilon)}(z-z')\hspace{5 cm}\nonumber\\
-\frac{1}{\epsilon_0^2}\int_{0}^{\infty} \rd z_1 D^{(\epsilon)}_{ij}(z-z_1)K_{jk}(\w)D^{(0)}_{kn}(z_1-z')\hspace{2cm}\nonumber\\
-\frac{1}{\epsilon_0^4}\int_0^\infty\rd z_1 \int_{-\infty}^0\rd z_2 D_{ij}^{(\epsilon)}(z-z_1)K_{jk}(\w)\hspace{2 cm}\nonumber\\
\times D_{kl}^{(0)}(z_1-z_2)K_{lm}(\w)D_{mn}(z_2,z')\;.\hspace{8mm}
\label{eqn:Combined}
\end{eqnarray}
In order to be able to solve Eq. (\ref{eqn:Combined}), we need to determine the displacement field propagator in the bulk anisotropic medium, that is, the quantity $D^{(\epsilon)}_{il}(z-z')$. To this end we write down Eq.~(\ref{eqn:IntegralAnisotropic}) with $g(\br_1)=1$ and Fourier transformed with respect to $\br-\br'$, 
\begin{equation}
\left[\delta_{il}-\frac{\delta_{ij}\bq^2-q_i q_j}{\w^2-\bq^2+i\eta}\frac{K_{jl}(\w)}{\epsilon_0}\right]D^{(\epsilon)}_{lk}(\bq,\w)=\epsilon_0\;\frac{\delta_{ik}\bq^2-q_i q_k}{\w^2-\bq^2+i\eta}
\end{equation}
where we have used the spectral representation of the free-space propagator from Eq. (\ref{eqn:FreePropagator}). Thus, finding $D^{(\epsilon)}_{lk}$ is just a question of finding the inverse of the matrix in the square brackets on the left-hand side. This matrix may be written as
\begin{equation}
\delta_{il}-\frac{\bq^2}{\w^2-\bq^2+i\eta}\sum_\lambda e_i^\lambda(\bq)e_j^\lambda(\bq)\frac{K_{jl}(\w)}{\epsilon_0}.\label{eqn:ToBeInverted}
\end{equation}
where we have used the completeness property of the transverse polarization vectors defined as
\begin{eqnarray}
\mathbf{e}^{\TE}(\bqp)&=&\frac{1}{|\bqp|}(q_y,-q_x,0),\nonumber\\
\mathbf{e}^{\TM}(\bqp,k_z)&=&\frac{1}{|\bqp|\w}(q_xk_z, q_yk_z,-\bqp^2).\label{eqn:PolVectors}
\end{eqnarray}
The products of the polarization vectors are projection operators onto the two orthogonal polarizations, so that their repeated application always gives the same polarization again. We have in particular 
\begin{equation}
e_i^\lambda(\bq)e_j^\lambda(\bq)K_{jl}(\w) e_l^\gamma(\bq)e_k^\gamma(\bq)= \delta^{\lambda\gamma} f^\lambda e_i^\lambda(\bq)e_k^\lambda(\bq) 
\label{eqn:NotMixTEwithTM}
\end{equation}
with 
\begin{eqnarray}
f^\TE &=& K_\parallel(\w),\nonumber\\
f^\TM &=&\dfrac{K_\perp(\w)\bqp^2+K_\parallel(\w)q_z^2}{\bq^2}.\nonumber
\end{eqnarray}
With that we are able to find the inverse of the matrix (\ref{eqn:ToBeInverted}) by writing it as sum over a geometrical series, according to
\begin{equation}
 ({\mathbb{1}}-\cal{O})^\mathrm{-1}={\mathbb{1}}+\cal{O}+\cal{O}^\mathrm{2}+\cdots
 \nonumber
\end{equation}
Equation (\ref{eqn:NotMixTEwithTM}) allows us to carry out the infinite summation, and we find
\begin{eqnarray}
D_{ij}(\br-\br';\w)&=&\frac{\epsilon_0\xi_\parallel(\w)}{(2\pi)^3}\sum_\lambda\int\rd^3 \bq\; Q^\lambda(\bq,\w)e_i^\lambda(\bq)e_j^\lambda(\bq)\nonumber\\
&\;&\times e^{i\bq\cdot(\br-\br')},\label{eqn:AnisotropicBulkProp}
\end{eqnarray}
with
\begin{eqnarray}
Q^\TE(\bq,\w) &=&\dfrac{\bq^2}{\xi_\parallel(\w)\w^2-\bq^2},\nonumber\\
Q^\TM(\bq,\w) &=&\dfrac{\bq^2}{\xi_\parallel(\w)\w^2-\left[\xi_\parallel(\w)/\xi_{\perp}(\w)\right]\bqp^2-q_z^2}.\nonumber
\end{eqnarray}
This result reproduces the bulk-medium propagator in an isotropic medium if we take $\xi_\parallel(\w)=\xi_\perp(\w)$. The transverse electric contribution to the propagator depends only on $\xi_\parallel(\w)$, which is not surprising because for this polarization the electric field in the direction perpendicular to the surface vanishes. Therefore, the TE part of the propagator is insensitive to the response of the material in the $\perp$ direction. The TM contributions to the propagator of course depend on both $\xi_\parallel(\w)$ and $\xi_\perp(\w)$. 

The quantity that enters the Eq.~(\ref{eqn:Combined}) is the propagator (\ref{eqn:AnisotropicBulkProp}) but Fourier transformed back from $q_z$  to $z-z'$. Carrying out the $q_z$  integration, one finds away from the point $z=z'$
\begin{eqnarray}
D_{ij}^{(\epsilon)}(z&-&z')=-i\epsilon_0\xi_\parallel(\w)\sum_\lambda\frac{(\bq_d^\lambda)^2}{2k_{zd}^\lambda}\nonumber\\
&\times &\left\{
\begin{array}{lr}
e_i^\lambda(k_{zd}^\lambda)e_j^\lambda(k_{zd}^\lambda)e^{ik_{zd}^\lambda(z-z')} & z>z'\\
e_i^\lambda(-k_{zd}^\lambda)e_j^\lambda(-k_{zd}^\lambda)e^{-ik_{zd}^\lambda(z-z')} & z<z'
\end{array} \right.\nonumber\\
&\;& \hspace{5 mm}\equiv \sum_\lambda D_{\lambda,ij}^{(\epsilon)}(z-z')\label{eqn:AnisotropicBulk}
\end{eqnarray}
with ${\bq}_d^\lambda=(\bqp, k_{zd}^\lambda)$ and
\begin{eqnarray}
k_{zd}^\TE &=&\sqrt{\xi_\parallel(\w)\w^2-\bqp^2},\nonumber\\
k_{zd}^\TM &=&\sqrt{\frac{\xi_\parallel(\w)}{\xi_\perp(\w)}}\sqrt{\xi_\perp(\w)\w^2-\bqp^2},\nonumber
\end{eqnarray}
where the square roots are taken such that their imaginary part is always positive. 
The medium affects the polarization vector of the TM mode which is given by
\begin{equation}
\mathbf{e}^{\TM}(\bqp,k^\TM_{zd})=\frac{1}{|{\bq}_d^\TM||\bqp|}(q_xk^\TM_{zd}, q_yk^\TM_{zd},-\bqp^2).\nonumber\\
\end{equation}
An equivalent representation for free-space propagator $D^{(0)}_{ij}(z-z')$ will also be needed in the following; it may be obtained simply by taking the limit $\xi_\sigma(\w)\rightarrow 1$ in Eq.~(\ref{eqn:AnisotropicBulk}).
We emphasize that formula (\ref{eqn:AnisotropicBulk}) is valid only away from the point $z=z'$, where additional singular terms would contribute. We also note that for the evaluation of the $z$ integrals in Eqs. (\ref{eqn:Halfspace1})--(\ref{eqn:Combined}) the polarization vectors need to be written in terms of differential operators acting on $e^{i\bqp\cdot(\brp-\brp')+ik^\lambda_{zd}|z-z'|}$.

Using the bulk and free-space propagators derived above, one can verify that for $z<0$ and $z'>0$ one has
\begin{eqnarray}
\frac{1}{\epsilon_0^4}\int_0^\infty\rd z_1 \int_{-\infty}^0\rd z_2 D_{ij}^{(\epsilon)}(z-z_1)K_{jk}(\w)\hspace{2 cm}\nonumber\\
\times D_{kl}^{(0)}(z_1-z_2)K_{lm}(\w)D_{mn}^{(\epsilon)}(z_2-z')\nonumber\\
=\sum_\lambda\frac{r_\lambda^2}{1-r_\lambda^2}D_{\lambda,in}^{(\epsilon)}(z-z').\hspace{5 mm}
\label{eq:stepIteration}
\end{eqnarray}
This features the Fresnel reflection coefficients at an anisotropic dielectric half-space,
\begin{equation}
r^\TE=\frac{k_z-k^\TE_{zd}}{k_z+k^{\TE}_{zd}},\;\;\;r^\TM=\frac{\xi_\parallel(\w)k_z-k^{\TM}_{zd}}{\xi_\parallel(\w)k_z+k^{\TM}_{zd}}.\label{eqn:AnisotropicRFresnels}
\end{equation}
with $k_z=\sqrt{\w^2-\bqp^2+i\eta}$, which is the $z$ component of the wave vector in vacuum. This has a small positive imaginary part which originates from the prescription of handling the poles in the free-space propagator in Eq.~(\ref{eqn:FreePropagator}) and serves to render the Fourier integrals meaningful.

Repeated use of Eq.~(\ref{eq:stepIteration}) facilitates the iteration of the integral equation (\ref{eqn:Combined}), yielding the expansion
\begin{eqnarray}
&\;&D_{\lambda,il}(z,z') =
\bigg[ 
D^{(\epsilon)}_{\lambda,ij}(z-z')
\hspace{3.5 cm}\nonumber\\
&\;&\hspace{5 mm}-\frac{1}{\epsilon_0^2}\int_0^\infty\rd z_1  D^{(\epsilon)}_{\lambda,ij}(z-z_1) 
K_{jk}(\w) D_{\lambda,kl}^{(0)}(z_1-z') 
\bigg]\nonumber\\
&\; & \hspace{5 mm} \times\left[1-\left(\dfrac{r_\lambda^2}{1-r_\lambda^2}\right)+\left(\dfrac{r_\lambda^2}{1-r_\lambda^2}\right)^2+\ldots\right]\hspace{.5 cm}\label{eqn:IntegralEqIterated2}
\end{eqnarray}
This is the photon propagator for the case $z<0$ and $z'>0$. Substituting the bulk and free-space propagators and summing the geometric series gives
\begin{eqnarray}
D_{ij}(z,z')&=&-\frac{i\epsilon_0\sqrt{\xi_\parallel(\w)\w^2}}{2k_z}\sum_\lambda |\bq_d^{\lambda}|t^\lambda e_i^\lambda(\bqp,-k_{zd}^\lambda)\nonumber\\
&\;&\times e_j^\lambda(\bqp,-k_{z}) e^{i k_z z'-i k_{zd}^{\lambda} z}\hspace{5 mm}\label{eqn:HSTransmitted}
\end{eqnarray}
with
\begin{equation}
t^\TE=\frac{2k_z}{k_z+k^{\TE}_{zd}},\;\;\;t^\TM=\frac{2\sqrt{\xi_\parallel(\w)}k_z}{\xi_\parallel(\w)k_z+k^{\TM}_{zd}}.
\end{equation}
The propagator for the case $z,z'>0$ can now be derived by applying the integral equation (\ref{eqn:Halfspace1}). It turns out to have the same form as that for the isotropic dielectric, derived in Ref.~\cite{Dispersive1}, but with the Fresnel reflection coefficients now those of Eq.~(\ref{eqn:AnisotropicRFresnels}) for an anisotropic half-space,
\begin{eqnarray}
D_{ij}(z,z')&=&D_{ij}^{(0)}(z-z')\nonumber\\
&-&\frac{i\epsilon_0\w^2}{2k_z}\sum_\lambda r^\lambda e_i^\lambda(\bqp,k_{z})e_j^\lambda(\bqp,-k_{z})\nonumber\\
&\;&\times e^{i k_z (z+z')}.\label{eqn:ReflectedPropagatorAnisotropic}
\end{eqnarray}
The photon propagator in coordinate space is obtained by inverse Fourier transform, i.e. by inverting Eq. (\ref{eqn:FourierTransform}).

\subsection{Photon propagator near an anisotropic dielectric slab}
With the photon propagator for a dielectric half-space determined, we are in position to work with integral equation (\ref{eqn:IntegralFinalAnisotropic}) and obtain the photon propagator for the slab geometry. Aiming to solve Eq.~(\ref{eqn:IntegralFinalAnisotropic}) by iteration, we evaluate 
\begin{eqnarray}
&\;&\frac{1}{\epsilon^4_0} \int_{L/2}^\infty \rd z_1 \int_{-\infty}^{-L/2}\rd z_2 D_{ik}^{(+)}(z,z_1)K_{kl}(\w) D_{lm}^{(-)}(z_1,z_2)\nonumber\\
&\;&\times
K_{mn}(\w)D^{(+)}_{nj}(z_2,z')=\sum_\lambda D_{\lambda,ij}^{(+)}(z,z')\left(r^\lambda e^{ik^\lambda_{zd}L}\right)^2\nonumber\\\label{eqn:IterationEnabler}
\end{eqnarray}
where we have restricted $z<-L/2$ and $z'>L/2$. To arrive at Eq.~(\ref{eqn:IterationEnabler}) we have used the result (\ref{eqn:HSTransmitted}), translated by $L/2$ for $D^{(-)}(z,z')$, and reflected and translated by $-L/2$ for $D^{(+)}(z,z')$. Equation (\ref{eqn:IterationEnabler}) shows that the action of the integral operator in Eq.~(\ref{eqn:IntegralFinalAnisotropic}) amounts to a simple multiplication by a factor of $\left(r^\lambda e^{ik^\lambda_{zd}L}\right)^2$. Therefore the iteration of Eq.~(\ref{eqn:IntegralFinalAnisotropic}) yields a geometric series that can be summed up to all orders,
\begin{eqnarray}
D_{\lambda,ij}(z,z')&=& \Bigg[D^{(+)}_{\lambda,ij}(z,z')-\frac{1}{\epsilon_0^2}\int_{L/2}^\infty \rd z_1 D^{(+)}_{\lambda,ik}(z,z_1)\nonumber\\
&\times & K_{kl}(\w)D^{(-)}_{\lambda,lj}(z_1,z')\Bigg]\frac{1}{1-\left(r^\lambda e^{ik^\lambda_{zd}L}\right)^2}\nonumber\\
\label{eqn:SlabExpansion}
\end{eqnarray}
This is the photon propagator in the slab geometry for the case $z<-L/2$ and $z'>L/2$, i.e.~it describes the transmission of photons across the slab. The $z_1$ integral can be evaluated, which allows us to write the final result in a more familiar form,
\begin{eqnarray}
D_{ij}(z,z')&=&-\frac{i\epsilon_0\w^2}{2k_z}\sum_\lambda T^\lambda e_i^\lambda(\bqp,-k_{z})e_j^\lambda(\bqp,-k_{z})\nonumber\\ &&\times e^{ik_z(z-z')}
\label{eqn:SlabTransmittedFinal}
\end{eqnarray}
where we have introduced the Fresnel transmission coefficients at an anisotropic slab,
\begin{equation}
T^\lambda=\frac{1-r_\lambda^2}{1-r^2_\lambda e^{2ik_{zd}^\lambda}} e^{i(k_{zd}^\lambda-k_z)L},
\end{equation}
and $r_\lambda$ are the same single-interface reflection coefficients as in Eq.~(\ref{eqn:AnisotropicRFresnels}) but we have written the polarization as a lower index for notational convenience. 
In order to obtain the propagator describing the reflection of photons from the slab, i.e.~for $z,z'>L/2$, we use the integral equation (\ref{eqn:IntegralAnisotropic2}). Substituting Eq.~(\ref{eqn:SlabTransmittedFinal}) into the right-hand side of Eq.~(\ref{eqn:IntegralAnisotropic2}) and using the results (\ref{eqn:HSTransmitted}) and (\ref{eqn:ReflectedPropagatorAnisotropic}), we find for the propagator in the region $z,z'>L/2$
\begin{eqnarray}
D_{ij}(z,z')&=&D_{ij}^{(0)}(z,z')\nonumber\\
&-&\frac{i\epsilon_0\w^2}{2k_z}\sum_\lambda R^\lambda e_i^\lambda(\bqp,k_{z})e_j^\lambda(\bqp,-k_{z})e^{ik_z(z+z')}\nonumber\\
\label{eqn:SlabReflectedFinal}
\end{eqnarray}
where $D_{ij}^{(0)}(z,z')$ is the (partially Fourier transformed) free-space photon propagator of Eq.~(\ref{eqn:FreePropagator}), and the Fresnel reflection coefficients $R^\lambda$ at an anisotropic slab are given by
\begin{equation}
R^\lambda=r^\lambda\frac{1-e^{2ik_{zd}^\lambda L}}{1-r^2_\lambda e^{2ik_{zd}^\lambda}}e^{-ik_zL}\; .
\label{eqn:ReflectionCoeffSlab}
\end{equation}
As one might have expected, the photon propagator near an anisotropic dielectric slab has exactly the same form as the one near an anisotropic dielectric half-space, Eq.~(\ref{eqn:ReflectedPropagatorAnisotropic}), except for the different Fresnel coefficients which account for each particular geometry of the dielectric.

Eqs. (\ref{eqn:SlabTransmittedFinal}) and (\ref{eqn:SlabReflectedFinal}) form a major result of this paper; they give the photon propagator for the quantized electromagnetic field near a slab of anisotropic dielectric material that is capable of absorbing radiation. Note that we have derived the propagator starting from a well defined microscopic model that is explicitly described by a Hamiltonian. We emphasize in particular that we have not made any reference to statistical physics for relating quantum propagators to classical Green's functions of the wave equation as is often done in the literature. Although the propagators we have derived describe the propagation of photons through inhomogeneous dielectric media, we did not have to explicitly invoke Maxwell's equations and continuity conditions following from them in the course of the derivation because this information is automatically included in the dynamics prescribed by the Halmiltonians. Also, note that throughout the derivation we have dealt only with physical electromagnetic fields and not with potentials that would have rendered our derivation gauge-dependent. For future reference we provide the complete result in coordinate space. For the source point to the right of the slab, $z'>L/2$, the photon propagator outside the slab reads
\begin{widetext}
\begin{eqnarray}
D_{ij}(\br,\br';\w) = \left\{ \begin{array}{ll}
\displaystyle
-\frac{i\epsilon_0}{(2\pi)^2}\int\rd^2 \bqp e^{i\bqp\cdot(\brp-\brp')}\frac{\w^2}{2k_z}\sum_{\lambda} 
T^\lambda e_i^\lambda(\bqp,-k_{z})e_j^\lambda(\bqp,-k_{z})e^{ik_z(z-z')}
& \mbox{for }z<-\frac{L}{2},\\
\displaystyle
D_{ij}^{(0)}(\br-\br';\w) 
-\frac{i\epsilon_0}{(2\pi)^2}\int\rd^2 \bqp e^{i\bqp\cdot(\brp-\brp')}\frac{\w^2}{2k_z}\sum_{\lambda} 
R^\lambda e_i^\lambda(\bqp,k_{z})e_j^\lambda(\bqp,-k_{z})e^{ik_z(z+z')}& \mbox{for }z>\frac{L}{2}.
\end{array}\right.
\label{eqn:SlabTotal}
\end{eqnarray}
\end{widetext}

\section{\label{sec:level3}Atom-dielectric interaction and energy-level shifts}
To study the interaction between an atom or molecule with an anisotropic dielectric slab, we choose a one-electron atom, either in its ground or in an excited state, interacting with a quantized electromagnetic field whose dynamics is governed by Eqs. (\ref{eqn:EMHam})--(\ref{eqn:FrequencyShifts}). We treat the atom using Schr\" odinger quantum mechanics but employ methods of second quantization. With the atomic energy levels known, the second-quantized Hamiltonian of the atom $H_A$ can be written as 
\begin{eqnarray}
H_{A}&=&\sum_n E_n c_n^\dagger c_n,\label{eqn:AtomicHamFinal}
\end{eqnarray}
where the creation and annihilation operators $c_i^\dagger$ and $c_j$ create an excitation in the atomic energy level $i$ and destroy one in $j$, respectively, and satisfy standard anti-commutation relations $\{c_i,c_j^\dagger\}=\delta_{ij}$. The atom is coupled to the field by a multipolar Hamiltonian, and the leading contribution to the interaction energy between atom and dielectric comes from the interaction between atom's electric dipole moment and the electromagnetic displacement field, which mediates the atom-surface interaction. In its second-quantized form the interaction Hamiltonian reads
\begin{eqnarray}
 H_{A-EM}&=&-\frac{e}{\epsilon_0}\sum_{ij}c_i^\dagger c_j\langle i |\boldsymbol{\rho}|j\rangle\cdot\mathbf{D}(\mathbf{R}),\label{eqn:AtomicHamCouplingFinal}
\end{eqnarray}
where $-e\langle i |\boldsymbol{\rho}|j\rangle$ are the matrix elements of atom's electric dipole moment and $\mathbf{R}$ is the position of the atom, which we always assume to be at least few Bohr radii away from the surface so as to be allowed to neglect direct wave function overlap between the atom and the dielectric medium and surface chemistry.
The energy-level shift in the atom arises as radiative correction to the self-energy of the atomic electron and may be calculated using the standard tools of quantum field theory, if one exploits the fact that the poles of a Feynman propagator correspond to the energy spectrum of the excitations \cite{Fetter}. Here we are after the atomic energy-levels which give rise to poles in the Fourier transform of the time-dependent propagator that describes the state of the atomic electron and in our formalism is given by
\begin{equation}
\mathcal{G}_{lm}(t,t')=-\frac{i}{\hbar}\left\langle \Omega|\T\left[c_l(t)c_m^\dagger(t')\right]|\Omega\right\rangle.\label{eqn:AtomicProp}
\end{equation}
Here $c_l(t)$ is a time-dependent operator in the Heisenberg picture, $|\Omega\rangle$ is the exact ground state of the system, and $\T$ is the time-ordering operator. 
The coupling (\ref{eqn:AtomicHamCouplingFinal}) between the atom and the dressed electromagnetic field changes the analytical structure of the atomic propagator (\ref{eqn:AtomicProp}) in the interacting as compared to the non-interacting case and shifts its poles and thereby the energy-levels of the atom. One part of this shift is the well-known Lamb shift, which is the same everywhere in space, and the other part depends on the distance of the atom to the surface because the interaction Hamiltonian $H_{A-EM}$, Eq.~(\ref{eqn:AtomicHamCouplingFinal}), contains the displacement field operator which takes into account the presence of the dielectric via its dynamics as prescribed by the Hamiltonians in Eqs. (\ref{eqn:EMHam})--(\ref{eqn:FrequencyShifts}) and is therefore governed by the photon propagators in Eqs.~(\ref{eqn:ReflectedPropagatorAnisotropic}) or (\ref{eqn:SlabTotal}) for the half-space or the slab, respectively.

The formulae for the energy-level shift of an atom interacting with the electromagnetic field as a functional of the photon propagator has been derived step by step in Ref.~\cite{Dispersive1}. The energy-level shift splits into two distinctive parts,
\begin{equation}
\Delta \mathcal{E}^{\rm ren}_i=\Delta \mathcal{E}_i+\Delta \mathcal{E}^{\star}_i\; ,\label{eqn:TotalShift}
\end{equation}
where the superscript 'ren' indicates that we are interested only in the renormalized energy-level shifts and have subtracted the free-space Lamb shift. In practice this renormalization procedure is very easily achieved thanks to the fact that the photon propagator outside the dielectric always splits into the free-space part $D^{(0)}_{ik}(\br-\br';\w)$ and the part that describes the reflection from the surface $D^{(r)}_{ik}(\br,\br';\w)$. Therefore, we just need to disregard the free-space part of the photon propagator and work only with the reflected part. With this in mind, the energy shift (\ref{eqn:TotalShift}) can be written as \cite{Dispersive1}
\begin{eqnarray}
\Delta \mathcal{E}_i&=&
\frac{1}{\pi\epsilon_0^2}\sum_{k,m}|\mu_{im}^k|^2\int_{0}^\infty\rd\xi\frac{\w_{mi}}{\xi^2+\w^2_{mi}}D^{(r)}_{kk}(\mathbf{R},\mathbf{R};i\xi), \hspace{.6 cm}\label{eqn:ShiftFinal}\\
\Delta \mathcal{E}^{\star}_i &=&\frac{1}{\epsilon_0^2}\sum_{k,m}|\mu_{im}^k|^2 D^{(r)}_{kk}(\mathbf{R},\mathbf{R};|\w_{mi}|)\theta(-\w_{mi}),\label{eqn:ShiftFinal1}
\end{eqnarray}
where $|\mu_{mi}^k|\equiv|\langle m|\mu^k|i\rangle|$ are the matrix elements of the $k$-th component of the electric dipole moment operator and $\w_{mi}=\w_m-\w_i$ is the frequency difference between the unperturbed atomic energy levels $|m\rangle$ and $|i\rangle$. The sums in Eqs.~(\ref{eqn:ShiftFinal}) and (\ref{eqn:ShiftFinal1}) run over the Cartesian components of the electric dipole moment operator $k=\{x,y,z\}$ and over all atomic states $|m\rangle$ excluding the state $|i\rangle$ that we are calculating the energy shift of. The sum over $m$ is in practice limited to one or a few states to which there are strong dipole transitions from the initial state $|i\rangle$. The shift $\Delta \mathcal{E}_i$ affects all atomic states but the contribution $\Delta \mathcal{E}^{\star}_i$ arises only if $|i\rangle$ is an excited state. Equations analogous to Eqs.~(\ref{eqn:ShiftFinal}) and (\ref{eqn:ShiftFinal1}) can also be derived by different methods e.g. by linear response theory \cite{McLachlan,Sipe} or by a phenomenological noise-current approach to macroscopic quantum electrodynamics \cite{Yoshi}. The shift $\Delta \mathcal{E}_i$ is always real because the photon propagator is real at complex frequencies. However, $\Delta \mathcal{E}_i^{\star}$ is complex and contains corrections to the spontaneous decay rates of excited states. One has
\begin{eqnarray}
\Delta E_i&=&{\rm Re}\left( \Delta \mathcal{E}_i^{\rm ren}\right)\nonumber\\
\Delta \Gamma_i&=&-\frac{2}{\hbar}{\rm Im}\left( \Delta \mathcal{E}_i^{\star}\right)\label{eqn:RatesDef}
\end{eqnarray}
where $\Delta E_i$ are the energy-level shifts and $\Delta \Gamma_i$ are the changes in decay rates. 

The photon propagators that we have derived in Sec.~\ref{sec:level2} are formally of the same form as the photon propagator that was derived in Ref.~\cite{Dispersive1} for the case of an isotropic dielectric half-space. The difference lies only in reflection coefficients. For this reason most of the formulae for the atomic energy-level shift derived in Ref.~\cite{Dispersive1} can be instantly generalized to the case of the anisotropic dielectric media studied here. First we are going to study the interaction between a neutral atom and an anisotropic dielectric half-space. Because the half-space geometry is relatively simple, we are able to study the effect of the anisotropy on the Casimir-Polder interaction in some detail. Then we go on to explore the physically more relevant case of a medium of finite thickness. But as the case of an anisotropic dielectric slab results in more complicated formulae for the energy-level shift, we will give fewer analytical and more numerical results than for the half-space.

\section{\label{sec:level4}Casimir-Polder shifts near an anisotropic half-space}
In this section we are going to derive a formula for the energy level shift of an atomic electron due to the presence of an anisotropic polarizable half-space. For this we focus on ground state shifts, but in Appendices \ref{sec:App1} and \ref{sec:App2} we also provide expressions for the shifts of excited energy levels and for the changes in the spontaneous decay rates due to the interaction with the half-space.

\subsection{Ground state shifts}
In order to obtain the general expression for the energy shift, we substitute the reflected part of the photon propagator (\ref{eqn:ReflectedPropagatorAnisotropic}) into Eq.~(\ref{eqn:ShiftFinal}) and obtain
\begin{eqnarray}
\Delta E_g=-\frac{1}{8\pi^2\epsilon_0}\sum_m\int_0^\infty\rd kk\int_{0}^\infty\rd\w\frac{\w_{mg}}{\w^2+\w_{mg}^2}\frac{e^{-2\sqrt{k^2+\w^2}\mathcal{Z}}}{\sqrt{k^2+\w^2}}\nonumber\\
\times\left\{\left[(k^2+\w^2)\bar{r}^\TM-\w^2\bar{r}^\TE\right]|\mu_{mg}^\parallel |^2+2k^2\bar{r}^\TM|\mu_{mg}^\perp|^2\right\},\nonumber\\\label{eqn:HalfSpaceShiftFinal}
\end{eqnarray}
with the abbreviation $|\mu_{mi}^\parallel |^2=|\mu_{mi}^x |^2+|\mu_{mi}^y |^2$. The reflection coefficients are defined in Eq.~(\ref{eqn:AnisotropicRFresnels}); in terms of the new variables used in Eq.~(\ref{eqn:HalfSpaceShiftFinal}) they read
\begin{eqnarray}
\bar{r}^\TE &=& \dfrac{\sqrt{\w^2+k^2}-\sqrt{\epsilon_\parallel\w^2+k^2}}{\sqrt{\w^2+k^2}+\sqrt{\epsilon_\parallel\w^2+k^2}},\label{eqn:BarredFresnelTE}\\
\bar{r}^\TM &=& \dfrac{\sqrt{\epsilon_\parallel \epsilon_\perp}\sqrt{\w^2+k^2}-\sqrt{\epsilon_\perp\w^2+k^2}}{\sqrt{\epsilon_\parallel\epsilon_\perp}\sqrt{\w^2+k^2}+\sqrt{\epsilon_\perp\w^2+k^2}}.\label{eqn:BarredFresnelTM}\;\;\;
\label{eqn:BarFresnelAnisotropic}
\end{eqnarray}
where all dielectric functions are evaluated at imaginary frequencies, that is, $\epsilon_\sigma=\epsilon_\sigma(i\w)$. Note that compared to Eq.~(\ref{eqn:AnisotropicRFresnels}) we have replaced the previously unnamed function $\xi_\sigma(\w)$ with the dielectric constant $\epsilon_\sigma(\w)$ because both functions coincide when evaluated at imaginary frequencies, cf.~Eq.~(\ref{eqn:Coincide}).

The expression for the energy shift in Eq.~(\ref{eqn:HalfSpaceShiftFinal}) is not necessarily the most convenient for asymptotic or numerical analysis. To find an alternative expression, we go to polar coordinates, $\w=x\w_{mg}\cos\phi,\;k=x\w_{mg }\sin\phi$, and then set $\cos\phi=y$. After short calculation we obtain
\begin{equation}
\Delta E_g=-\frac{1}{8\pi^2\epsilon_0\mathcal{Z}^4}\sum_m\frac{1}{\w_{mg}}\left[F^\parallel |\mu_{mg}^\parallel |^2+F^\perp|\mu_{mg}^\perp |^2\right]\label{eqn:HalfSpaceShiftFinal2Anisotropic}
\end{equation}
with $|\mu_{mg}^\parallel |^2=|\mu_{mg}^x |^2+|\mu_{mg}^y |^2$ and
\begin{eqnarray}
F^\parallel &=&\int_0^\infty\rd xx^3\int_{0}^1\rd y \frac{(\mathcal{Z}\w_{mg})^4}{1+x^2y^2}
\left(\widetilde{r}^\TM-y^2\widetilde{r}^\TE\right)e^{-2\w_{mg}\mathcal{Z}x},\nonumber\\\label{eqn:FPara}\\\nonumber
F^\perp &=&\int_0^\infty\rd xx^3\int_{0}^1\rd y \frac{(\mathcal{Z}\w_{mg})^4}{1+x^2y^2}\left(1-y^2\right)2\widetilde{r}^\TM e^{-2\w_{mg}\mathcal{Z}x}.\\\label{eqn:FPerp}
\end{eqnarray}
In these variables the reflection coefficients read
\begin{eqnarray}
\widetilde{r}^\TE&=&\frac{1-\sqrt{y^2[\epsilon_\parallel-1]+1}}{1+\sqrt{y^2[\epsilon_\parallel-1]+1}},\nonumber\\
\widetilde{r}^\TM&=&\frac{\sqrt{\epsilon_\parallel\epsilon_\perp}-\sqrt{y^2(\epsilon_\perp-1)+1}}{
\sqrt{\epsilon_\parallel\epsilon_\perp}+\sqrt{y^2(\epsilon_\perp-1)+1}},\hspace{5mm}\label{eqn:TildedFresnelAnisotropc}
\end{eqnarray}
where $\epsilon_\sigma=\epsilon_\sigma(ixy\w_{mg})$, that is, for example,
\begin{equation}
\epsilon_\parallel(ixy\w_{mg})=1+
\frac{(\w_{\rm P\parallel}/\w_{mg})^2}
{x^2y^2+(\w_{\rm T\parallel}/\w_{mg})^2+2xy(\gamma_\parallel /\w_{mg})}.\nonumber
\end{equation}
Thus all frequencies entering Eq.~(\ref{eqn:HalfSpaceShiftFinal2Anisotropic}) are measured in units of the atomic transition frequency $\w_{mg}$ and the argument of the exponential $\mathcal{Z}\w_{mg}=2\pi\mathcal{Z}/\lambda_{mg}$ is the atom-surface distance measured in units of atomic transition's wavelength $\lambda_{mg}$. Note that the sum over $m$ in Eq.~(\ref{eqn:HalfSpaceShiftFinal2Anisotropic}) also contains the dipole matrix elements, whence in practice it is dominated by just one or a few states with strong dipole transitions from the initial (ground) state $|g\rangle$. We shall refer to $\w_{mg}$ as the frequency of a typical atomic transition.

The expression for the energy shift given in Eq.~(\ref{eqn:HalfSpaceShiftFinal2Anisotropic}) (or equivalently in Eq.~(\ref{eqn:HalfSpaceShiftFinal})) is general in the sense that $\epsilon_\parallel(\w)$ and $\epsilon_\perp(\w)$ are arbitrary as long as they can be reproduced by the oscillator model that we have started with. Note in particular that this includes dielectrics with more than one absorption line, as discussed in Refs.~\cite{Janowicz,Dispersive1}. In the following we would like to consider the asymptotic behaviour of Eqs.~(\ref{eqn:HalfSpaceShiftFinal}) and (\ref{eqn:HalfSpaceShiftFinal2Anisotropic}) in a few physically important limits.

\subsubsection{Nonretarded limit\label{sec:Lev4nonret}}
First we consider the non-retarded (or electrostatic) limit where electromagnetic interactions are instantaneous and the speed of light is infinite, 
$c\rightarrow \infty$. Having worked hitherto in natural units where $c=1$, we recall that for the frequency $\w$ to have the same dimensions as the wave vector $k$ it needs to be multiplied by $1/c$. Restoring the missing factors of $1/c$ in the appropriate places and taking the limit $c\rightarrow\infty$, we can then perform the $k$ integration in Eq.~(\ref{eqn:HalfSpaceShiftFinal}), and the final result can be cast in the form
\begin{eqnarray}
\Delta E^{\rm nonret}_g \approx -\frac{1}{32\pi^2\epsilon_0\mathcal{Z}^3}\sum_m \left(|\mu_{mg}^\parallel|^2+2|\mu_{mg}^\perp|^2\right)\nonumber\\
 \int_0^\infty\rd \w\;\frac{\w_{mg}}{\w^2+\w_{mg}^2}\;\frac{\sqrt{\epsilon_\parallel(i\w)\epsilon_\perp(i\w)}-1}{\sqrt{\epsilon_\parallel(i\w)\epsilon_\perp(i\w)}+1}.\hspace{.5 cm}\label{eqn:NonRet}
\end{eqnarray}
Thus in the nonretarded regime, the ground-state shift caused by the anisotropic medium depends on the distance $\mathcal{Z}$ of the atom from the surface as $\mathcal{Z}^{-3}$ (as it also would for an isotropic dielectric half-space or a perfectly reflecting surface because in all these cases the underlying interaction is an electrostatic dipole-dipole interaction in the same geometry). The coefficient of $Z^{-3}$ arises as an integral over the atomic polarizability along the imaginary frequency axis with a factor
 \begin{equation}
 \frac{\sqrt{\epsilon_\parallel(i\w)\epsilon_\perp(i\w)}-1}{\sqrt{\epsilon_\parallel(i\w)\epsilon_\perp(i\w)}+1}
 \end{equation}
which is a generalization of the electrostatic image factor that would have arisen if one were to determine the Green's function of the Poisson equation for an anisotropic dielectric half-space. We note that setting $\epsilon_\parallel=\epsilon_\perp$ and neglecting the damping leads to the result reported in Ref.~\cite{BabikerBarton} where the shifts were calculated by explicitly quantizing the Maxwell field coupled to a half-space filled with a plasma. Formula (\ref{eqn:NonRet}) is also valid for dielectrics or, in fact, for any material whose dielectric function can be described by the Lorentz-Drude oscillator model. In other words, the electrostatic limit $c\rightarrow\infty$ does not interfere with the limit $\w_{\rm T}\rightarrow 0$ which corresponds to the {dielectric $\rightarrow$ conductor} limit. This is not always the case and in general the order in which various limits are taken matters and great care must be taken in asymptotic expansions of energy-level shifts \cite{massshift}.

The limit $c\rightarrow \infty$ corresponds to the atom being very close to the surface, by which we mean that $\mathcal{Z}\w_{mg}\ll 1$ where $\w_{mg}$ is the frequency of the dominant dipole transition in the atom. The parameter $2\mathcal{Z}\w_{mg}/c$ (if we restore the factor of 1/c) serves as a measure of how fast the atom evolves as compared to the time taken by a photon for one round trip between the atom and the surface. If $\mathcal{Z}\w_{mg}\ll 1$, the atom does not evolve appreciably while the photon travels to the surface and back, which is equivalent to taking the speed of light as infinite. 

\subsubsection{Retarded limit}
In the opposite regime, if $\mathcal{Z}\w_{mg}\gg 1$, retardation is crucially important and well-known to have a significant impact on atom-surface interactions \cite{CP}.
The retarded limit $\mathcal{Z}\w_{mg}\rightarrow\infty$ is not interchangeable with either the {dielectric $\rightarrow$ conductor} limit, $\w_{\rm T}\rightarrow 0$, nor with the limit of no damping, $\gamma\rightarrow 0$. Therefore, when working out the asymptotic expansion of the energy shift in the retarded limit, we need to specify a definite form of the electromagnetic response of the material from the outset. Let us first examine the case of an anisotropic conductor with losses, where we take
\begin{eqnarray}
\epsilon_\perp(\w)&=&1+\frac{i}{\w}\frac{\omega_{\rm P\perp}^2}{2\gamma_\perp-i\w}
\equiv 1+i\frac{\sigma_\perp(\w)}{\w}\;,\nonumber\\
\epsilon_\parallel(\w)&=&1+\frac{i}{\w}\frac{\omega_{\rm P\parallel}^2}{2\gamma_\parallel-i\w}
\equiv 1+i\frac{\sigma_\parallel(\w)}{\w}\;, \label{eqn:Conductivity}
\end{eqnarray}
where $\sigma$ is the usual Drude conductivity. For $\mathcal{Z}\w_{mg}\gg1$ the $x$ integral in Eq.~(\ref{eqn:HalfSpaceShiftFinal2Anisotropic}) is strongly damped by the exponential. Then, according to Watson's lemma \cite{Bender}, we obtain an asymptotic expansion of the integral by keeping the exponential, expanding the remaining part of the integrand in a power series around $x=0^+$, and carrying out the $x$ and $y$ integrations analytically. We find that the asymptotic expansion of the $F^\rho$ coefficients entering Eq. (\ref{eqn:HalfSpaceShiftFinal2Anisotropic}) can be written as
\begin{eqnarray}
F^{\parallel,\perp}\approx\frac{1}{2}\left[1-\frac{C^{\parallel,\perp}_{9/2}}{\sqrt{\sigma_\parallel(0)\mathcal{Z}}}+\frac{C^{\parallel,\perp}_{5}}{\sigma_\parallel(0)\mathcal{Z}}-\ldots\right]
,\label{eqn:retardedHSAnisotropic}
\end{eqnarray}
with $\sigma_{\parallel}(0)=\w_{\rm P\parallel}^2/2\gamma$ and 
\begin{equation}
C^\parallel_{9/2}=\frac{21}{16}\sqrt{\frac{\pi}{2}},\;\;\;
C^\perp_{9/2}=\frac{7}{12}\sqrt{\frac{\pi}{2}},\;\;\;
C^\parallel_{5}=\frac{9}{4},\;\;\;
C^\perp_{5}=\frac{1}{2}.\nonumber
\label{eqn:CCoefficients}
\end{equation}
The first term in Eq.~(\ref{eqn:retardedHSAnisotropic}) reproduces the result of Casimir and Polder for a perfectly reflecting mirror \cite{CP}. This is what one would expect because in the retarded regime it is the static response of both the atom and the material that matters the most \cite{McLachlan}. In our model of an anisotropic conductor we have perfect screening in the static limit, i.e. $\epsilon(\w)\rightarrow\infty$ for $\w \rightarrow 0$, just like for an isotropic material. That is why the leading terms of the asymptotic expansion (\ref{eqn:retardedHSAnisotropic}) do not depend on $\sigma_\perp$ and are exactly the same as for an isotropic medium with $\sigma_\parallel=\sigma_\perp$. In other words, to leading-order, the anisotropy of the conductor does not matter in the retarded regime.

It is worth noting that the expansion (\ref{eqn:retardedHSAnisotropic}) is not compatible with the vacuum limit $\w_{\rm P \parallel}\rightarrow 0$. In this limit we would expect the shifts to vanish, but this is clearly not the case for formula (\ref{eqn:retardedHSAnisotropic}) as it contains the perfect-reflector terms which are independent of $\sigma_\parallel(0)$. Not only that; the subsequent terms in the expansion even diverge in the vacuum limit. This is because we have no right to expect the retarded limit $\mathcal{Z}\w_{mg}\gg 1$ to be compatible with the low-conductivity limit $\mathcal{Z}\w_{\rm P\parallel}\ll 1$, as has been noted before \cite{BabikerBarton}. 

Note also that in the limit $\gamma_\parallel\rightarrow 0$ of no damping only the leading-order term survives. One might be tempted to think that in the absence of damping any corrections to the perfect-reflector behaviour of the energy shift are of higher order than $(\mathcal{Z}\w_{mg})^{-5}$, but this is not correct. The no-damping limit $\gamma_\parallel\rightarrow 0$ is not interchangeable with the retarded limit $\mathcal{Z}\w_{mg}\rightarrow \infty$, and this becomes more apparent if we derive the next term of the expansion (\ref{eqn:retardedHSAnisotropic}), which turns out to diverge in the limit of no damping. Therefore, the undamped case $\gamma_\parallel \rightarrow 0$ must be treated separately by first setting $\gamma_\parallel=0$ in Eq.~(\ref{eqn:Conductivity}) and then repeating the steps leading to Eq.~(\ref{eqn:retardedHSAnisotropic}). Then we get for an undamped conductor
\begin{eqnarray}
F^{\parallel,\perp}\approx\frac{1}{2}\left(1-\frac{C^{\parallel,\perp}_{5}}{\mathcal{Z}\w_{\rm P\parallel}}+\ldots\right)
,\label{eqn:retardedHSAnisotropic2}
\end{eqnarray}
with $C_5^\perp=4/5$ and $C_5^\parallel=2$. 

\begin{figure}[ht]
\includegraphics[width=8.5 cm, height=6 cm]{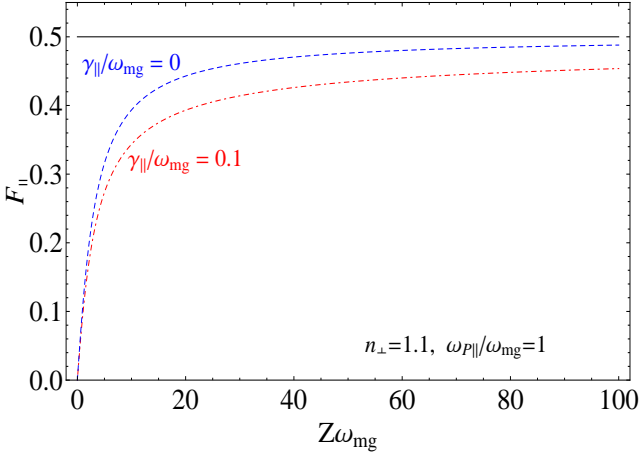}% Here is how to import EPS art
\caption{\label{fig:RoleOfDamping} (Color online) The dimensionless $F_\parallel$ of Eq.~(\ref{eqn:FPara}) that enters the shift (\ref{eqn:HalfSpaceShiftFinal2Anisotropic}) for an electromagnetic response of the form (\ref{eqn:Conductivity2}), as a function of the dimensionless parameter $Z\w_{mg}$, i.e.~the distance from the surface measured in atomic transition wavelengths. The solid straight line shows the retarded limit for the perfect reflector, which is the leading-order term in the asymptotic expansion of $F_\parallel$, Eqs.~(\ref{eqn:retardedHSAnisotropic3}) and (\ref{eqn:retardedHSAnisotropic4}). In the presence of damping (red, dot-dashed) i.e. for non-zero  $\gamma_\parallel$, even for relatively large values of $\mathcal{Z}\w_{mg}$, the corrections to the leading-order behaviour are still appreciable, especially for poor conductors. Without damping (blue, dashed) they are much smaller.}
\end{figure}

Since the asymptotic expansion (\ref{eqn:retardedHSAnisotropic}) applies only to anisotropic conductors, i.e.~only to materials whose electromagnetic response is described by Eq.~(\ref{eqn:Conductivity}), let us consider for comparison a case where the material behaves as a non-dispersive dielectric in the direction perpendicular to the interface and as a lossy conductor in the direction parallel to the surface. In other words, we take
\begin{eqnarray}
\epsilon_\parallel(\w)=1+i\frac{\sigma_\parallel(\w)}{\w},\;\;\;
\epsilon_\perp(\w)=n_\perp^2.
\label{eqn:Conductivity2}
\end{eqnarray}
Then, repeating the steps described in the paragraph below Eq.~(\ref{eqn:Conductivity}), we find that in the retarded regime the coefficients $F^\rho$ entering Eq.~(\ref{eqn:HalfSpaceShiftFinal2Anisotropic}) are given by
\begin{eqnarray}
F^{\parallel,\perp}\approx\frac{1}{2}\left[1-\frac{C^{\parallel,\perp}_{9/2}}{\sqrt{\sigma_\parallel(0)\mathcal{Z}}}+\frac{C^{\parallel,\perp}_{5}}{\sigma_\parallel(0)\mathcal{Z}}-\ldots\right]
,\label{eqn:retardedHSAnisotropic3}
\end{eqnarray}
with 
\begin{eqnarray}
C^\parallel_{9/2}&=&\sqrt{\frac{\pi}{2}}\left[\frac{21}{64}+\frac{35}{64}\frac{1}{n_\perp}\;_2{\rm F}_1\left(-\frac{1}{2},\frac{3}{4},\frac{7}{4};1-n^2_\perp\right)\right],\nonumber\\
C^\perp_{9/2}&=&\sqrt{\frac{\pi}{2}}\frac{7}{8}\frac{1}{n^2_\perp-1}\left[\frac{n^2_\perp}{3}-\frac{1}{2}+\frac{1}{2}\;_2{\rm F}_1\left(\frac{5}{4},1,\frac{7}{4};1-n^2_\perp\right)\right.\nonumber\\
&-&\left.\frac{1}{3n_\perp}\;_2{\rm F}_1\left(\frac{1}{2},\frac{3}{4},\frac{7}{4};1-n^2_\perp\right)\right],\nonumber\\
C^\parallel_{5}&=&\frac{3}{8}\frac{3n_\perp^2+1}{n^2_\perp}\;,\nonumber\\
C^\perp_{5}&=&\frac{n^2_\perp+2}{4n^2_\perp}\;,\nonumber
\end{eqnarray}
As one can see, the coefficients $C^{\parallel,\perp}_{9/2}$ are rather complicated hypergeometric functions of the static refractive index $n_\perp$ and we have listed them here for completeness only. We emphasize that, for the same reasons as before, the no-damping limit $\gamma_\parallel\rightarrow 0$ is not permitted in the expansion (\ref{eqn:retardedHSAnisotropic3}). To study the case of no damping we must set $\gamma_\parallel=0$ in Eq.~(\ref{eqn:Conductivity2}) and repeat the asymptotic analysis. Then, we get 
\begin{eqnarray}
F^{\parallel,\perp}\approx\frac{1}{2}\left(1-\frac{C^{\parallel,\perp}_{5}}{\mathcal{Z}\w_{\rm P\parallel}}+\ldots\right)
,\label{eqn:retardedHSAnisotropic4}
\end{eqnarray}
with
\begin{eqnarray}
C^\parallel_5 &= &1+\frac{1}{2n_\perp(n_\perp+1)}\;,\nonumber\\
C^\perp_5 &= &\frac{2n_\perp^3+4n_\perp^2+6n_\perp+3}{5n_\perp\left(n_\perp+1\right)^2}\;.\nonumber
\end{eqnarray}
Interestingly, the expansions (\ref{eqn:retardedHSAnisotropic}) and (\ref{eqn:retardedHSAnisotropic3}) derived with damping, $\gamma_\parallel\neq 0$, contain fractional powers of $\mathcal{Z}$ which are not present in the absence of damping, cf.~Eqs.~(\ref{eqn:retardedHSAnisotropic2}) and (\ref{eqn:retardedHSAnisotropic4}). This can be traced back to the behaviour of $\sigma(\w)$ at $\w=0$, which varies depending on the presence (or absence) of dissipation in the material. Note that because of the presence of the fractional powers of $\mathcal{Z}$ in the asymptotic expansions (\ref{eqn:retardedHSAnisotropic}) and (\ref{eqn:retardedHSAnisotropic3}) the significance of next-to-leading order terms is much greater there than in expansions (\ref{eqn:retardedHSAnisotropic2}) and (\ref{eqn:retardedHSAnisotropic4}). Because these next-to-leading order terms depend on parameters describing losses in the material, we conclude that damping plays an important role in the Casimir-Polder interaction between an atom and a lossy conductor. This is in contrast to the case of an atom interacting with an absorptive insulator where even in the presence of damping there are no fractional powers of $\mathcal{Z}$ present in the asymptotic expansion of the energy shift in the retarded regime \cite{Dispersive1}. For lossy dielectrics the leading-order behaviour of the energy shift is the same as for a non-absorptive dielectric and the next-to-leading order term, which depends on the absorption parameter, is indeed a small correction. In Fig. \ref{fig:RoleOfDamping} we illustrate numerically the impact of damping for the case of a material that conducts parallel to the surface but insulates perpendicular to it. 
%%%%%%%%%%%%%%%%%%%%%%%%%%%%%%%%%%%%%
%%%%%%%%%%%%%%%%%%%%%%%%%%%%%%%%%%%%%%%%

Comparing Eqs. (\ref{eqn:retardedHSAnisotropic}) and (\ref{eqn:retardedHSAnisotropic3}) together with Eqs. (\ref{eqn:retardedHSAnisotropic2}) and (\ref{eqn:retardedHSAnisotropic4}) we see that in the case when the response of the material in the $z$ direction is taken to mimic a dielectric rather than a conductor, the quantities pertaining to $\epsilon_\perp$ appear earlier in the asymptotic series. Note that if $\epsilon_\parallel$ and $\epsilon_\perp$ both describe dielectrics, then already the leading-order term of the asymptotic expansion depends on both $n_\parallel$ and $n_\perp$, where $n_{\perp,\parallel}$ are the static refractive indices i.e. $n^2_{\perp,\parallel}=\epsilon_{\perp,\parallel}(0)$. Thus, we conclude that in the retarded limit the anisotropy of the material is important in the case of an atom interacting with a dielectric but not so much if it is a conductor.

\subsubsection{Case $\w_{\rm P\parallel}\gg\w_{\rm P\perp}$}
We now turn to study in greater detail the scenario of an atom interacting with a material whose conductivity in the direction parallel to the interface far exceeds that in the direction perpendicular to the interface. In particular, we are interested in the behaviour of the energy shift in the non-retarded regime, where the impact of the material's anisotropy is greatest. We assume $\w_{\rm P\parallel}\gg\w_{\rm P\perp}$ and thereby effectively set $\sigma_\perp=0$. We also neglect damping, that is, we take $\gamma_\parallel=0$. 
With these simplifications the shift is still given by the Eq.~(\ref{eqn:HalfSpaceShiftFinal2Anisotropic}) but the reflection coefficients are now significantly simpler
\begin{eqnarray}
\widetilde{r}_{\TE}&=&\frac{2x}{\wP^2}\left(\sqrt{\wP^2+x^2}-x\right)-1,\\
\widetilde{r}_{\TM}&=&\frac{2xy}{\wP^2}\left(xy-\sqrt{\wP^2+x^2y^2}\right)+1,
\end{eqnarray}
where we have abbreviated $\wP=\w_{\rm P\parallel}/\w_{mg}$. The $y$ integral is now  elementary and can be calculated analytically, leading to
\begin{eqnarray}
F^\parallel &=&(\mathcal{Z}\w_{mg})^4\int_0^\infty\rd x\; e^{-2\mathcal{Z}\w_{mg}x}\Bigg\{(x^2-1)\arctan(x)+x\nonumber\\
&+&\frac{2x}{\wP^2}\left(2x-\sqrt{\wP^2+x^2}\right)\left[x-\arctan(x)\right]\nonumber\\
&-&\frac{2x^2}{\wP^2}\Bigg[\sqrt{\wP^2+x^2}-\wP-\sqrt{1-\wP^2}\nonumber\\
&\times &\left(\arctan\frac{\sqrt{\wP^2+x^2}}{\sqrt{1-\wP^2}}-\arctan\frac{\wP}{\sqrt{1-\wP^2}}\right)\Bigg]\Bigg\},\nonumber\\
F^\perp &=&\frac{4}{\wP^2}(\mathcal{Z}\w_{mg})^4\int_0^\infty\rd x\; e^{-2\mathcal{Z}\w_{mg}x}\Bigg\{\frac{2x^3}{3}\label{eqn:FFunctions2}\\
&+&\frac{1}{3}\left[(x^2+\wP^2)^{3/2}-\wP^3 \right]+(x^2+1)\sqrt{1-\wP^2}\nonumber\\
&\times &\left(\arctan\frac{\sqrt{x^2+\wP^2}}{\sqrt{1-\wP^2}}-\arctan\frac{\wP}{\sqrt{1-\wP^2}}\right)\nonumber\\
&-&(x^2+1)\left(\sqrt{\wP^2+x^2}-\wP\right)+\left(\frac{\wP^2}{2}-1\right)\nonumber\\
&\times & \left[(x^2+1)\arctan(x)-x\right]\Bigg\},\nonumber
\end{eqnarray}
where the coefficients $F^\rho$ are understood to enter Eq.~(\ref{eqn:HalfSpaceShiftFinal2Anisotropic}) to give the energy shift. The expressions in Eq.~(\ref{eqn:FFunctions2}) may look formidable but in fact they are given entirely in terms of elementary functions. More importantly, the coefficients $F^\rho$ are given as fast-converging one-dimensional integrals \cite{fn2} that can almost effortlessly be computed numerically, and the subtleties of the asymptotic analysis of two-dimensional integrals need not concern us here \cite{EberleinSiklois}. Note that terms containing $\sqrt{1-\wP^2}$ can in principle become imaginary for $\wP>1$, although in practice this is unlikely to happen. As a typical example let us consider graphite with a plasma frequency of the order of $\w_{\rm P}\approx 0.7 \times 10^{15}$ Hz \cite{Phillip} and a rubidium ${}^{87}$Rb atom  which has its strongest dipole transition ($5^2{\rm S}_{1/2}\rightarrow 5^2{\rm P}_{3/2}$) at $\w_{mg}\approx 2.4\times 10^{15}$ Hz \cite{Safronova}. Then, we have $\wP=\w_{\rm P}/\w_{mg}\approx 0.29$. However, even if $\wP>1$ the expressions in Eq.~(\ref{eqn:FFunctions2}) actually remain real, but feature inverse hyperbolic functions through $\arctan(iz)=i\, \mbox{arctanh}(z)$. 

\begin{figure}[ht]
\includegraphics[width=8.5 cm, height=6 cm]{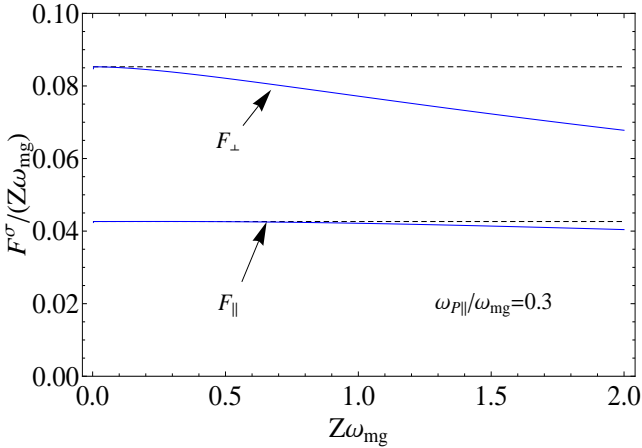}
\caption{\label{fig:hsnums2} (Color online) For small $\mathcal{Z}\w_{mg}$ we choose to plot $F^\rho/(\mathcal{Z}\w_{mg})$ which according to Eq.~(\ref{eqn:NRFperp}) approaches a constant as $\mathcal{Z}\w_{mg}\rightarrow 0$, so that the energy shift in Eq.~(\ref{eqn:HalfSpaceShiftFinal2Anisotropic}) behaves as $\mathcal{Z}^{-3}$ in the nonretarded limit. The solid lines represent the results of the exact numerical integration of Eq.~(\ref{eqn:FFunctions2}) whereas the dashed lines come from the corresponding approximate formulae in Eqs.~(\ref{eqn:NRFpara}, \ref{eqn:NRFperp}).}
\end{figure}

Let us now consider the nonretarded limit of Eq.~(\ref{eqn:FFunctions2}). In order to obtain the asymptotic expansion we note that in the limit $\mathcal{Z}\w_{mg}\rightarrow 0$ the exponential in the integrand approaches unity and the integral diverges. Thus we replace the integrand by its large-$x$ behaviour and retain only the positive powers of $x$. Then, carrying out the integrations we obtain
\begin{eqnarray}
F^{\parallel}&\approx &\mathcal{Z}\w_{mg}\left(\frac{\pi}{8}-\frac{1}{2\wP}-\frac{\pi}{4\wP^2}\right.\nonumber\\
&-&\left.\frac{\sqrt{1-\wP^2}}{2\wP^2}{\rm arccot}\frac{\wP}{\sqrt{1-\wP^2}}\right),\label{eqn:NRFpara}\\
F^{\perp}&\approx &\mathcal{Z}\w_{mg}\left(\frac{\pi}{4}-\frac{1}{\wP}-\frac{\pi}{2\wP^2}\right.\nonumber\\
&-&\left.\frac{\sqrt{1-\wP^2}}{\wP^2}{\rm arccot}\frac{\wP}{\sqrt{1-\wP^2}}\right)\label{eqn:NRFperp}
%&+&(\mathcal{Z}\w_{mg})^2\left(\frac{\pi}{2}-\frac{\pi}{\wP^2}+\frac{2}{\wP}-\frac{2}{3}\wP\right.\nonumber\\
%&-&\left.\frac{2\sqrt{1-\wP^2}}{2\wP^2}{\rm arccot}\frac{\wP}{\sqrt{1-\wP^2}}\right).
\end{eqnarray}
Inserted into Eq.~(\ref{eqn:HalfSpaceShiftFinal2Anisotropic}) these results show that to leading order the shifts depend on distance as $\mathcal{Z}^{-3}$, as one would expect for electrostatic interactions. We note that the leading-order contributions each contain a $\wP$-independent term, which is the only one surviving the limit $\tilde{\w}_{\rm P \parallel}\rightarrow\infty$ where it yields the perfect reflector result, as naively expected. Thus, one could see the results (\ref{eqn:NRFpara}, \ref{eqn:NRFperp}) as those for a perfect reflector but amended by additional material-dependent corrections that are of the same order in the parameter $\mathcal{Z}\w_{mg}$. Note also that both $F^\parallel$ and $F^{\perp}$ vanish in the vacuum limit $\wP\rightarrow 0$, as they should. In Fig.~\ref{fig:hsnums2} we compare the asymptotic expansions in Eqs.~(\ref{eqn:NRFpara}, \ref{eqn:NRFperp}) with the results of the exact numerical integration of Eq.~(\ref{eqn:FFunctions2}).
%%%%%%%%%%%%%%%%%%%%%%%%%%%%%%%%%%%%%
%%%%%%%%%%%%%%%%%%%%%%%%%%%%%%%%%%%%%%%%

\begin{figure}[ht]
\includegraphics[width=8.5 cm, height=6 cm]{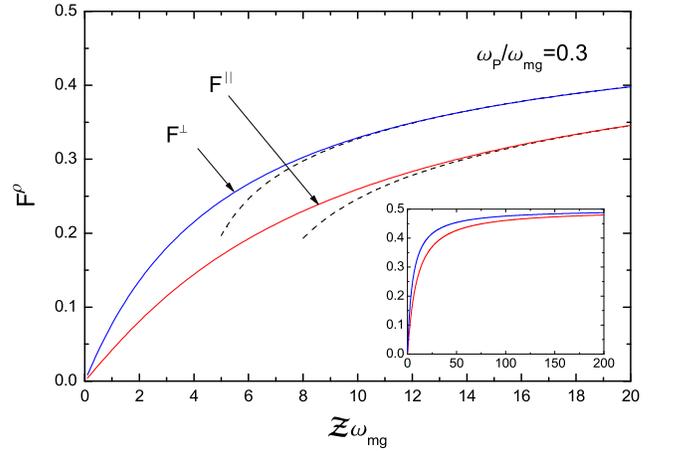}
\caption{\label{fig:hsnums3} (Color online)  
For large $\mathcal{Z}\w_{mg}$, the coefficients $F^\rho$ according to Eq.~(\ref{eqn:NRFperp}) approach constants, which implies that the energy shift in Eq.~(\ref{eqn:HalfSpaceShiftFinal2Anisotropic}) behaves as $\mathcal{Z}^{-4}$ in the retarded regime. The solid lines represent results of the exact numerical integration of Eq.~(\ref{eqn:FFunctions2}) whereas the dashed lines have been calculated from the corresponding approximate formulae in Eqs.~(\ref{eqn:RetFPara}, \ref{eqn:RetFPerp}). Note that the corrections to the leading-order terms may be significant for small values of $\w_P/\w_{mg}$.}
\end{figure}

In the retarded limit the asymptotic expansion of Eq.~(\ref{eqn:FFunctions2}) is obtained by observing that the integrand is again strongly damped in the limit $\mathcal{Z}\w_{mg}\rightarrow\infty$. Thus, we separate out the exponential and replace the remaining part of the integrand with its small-$x$ behaviour. Integrating the resulting expression term by term we arrive at
\begin{eqnarray}
F^\parallel &\approx &\frac{1}{2}-\frac{5}{4\wP}\frac{1}{\mathcal{Z}\w_{mg}}+\frac{5-2\wP^2}{2\wP^2}\frac{1}{(\mathcal{Z}\w_{mg})^2}\nonumber\\
&+&\frac{162\wP^2-105}{32\wP^3}\frac{1}{(\mathcal{Z}\w_{mg})^3}+\ldots,\label{eqn:RetFPara}\\
F^\perp &\approx &\frac{1}{2}-\frac{3}{4\wP}\frac{1}{\mathcal{Z}\w_{mg}}+\frac{2-\wP^2}{2\wP^2}\frac{1}{(\mathcal{Z}\w_{mg})^2}\nonumber\\
&+&\frac{30\wP^2-15}{16\wP^3}\frac{1}{(\mathcal{Z}\w_{mg})^3}+\ldots .\label{eqn:RetFPerp}
\end{eqnarray}
This expansion of course coincides with the one in Eq.~(\ref{eqn:retardedHSAnisotropic4}) in the limit $n_\perp\rightarrow 1$. A comparison of asymptotic and numerical results in the retarded limit is shown in Fig.~\ref{fig:hsnums3}.
%%%%%%%%%%%%%%%%%%%%%%%%%%%%%%%%%%%%%
%%%%%%%%%%%%%%%%%%%%%%%%%%%%%%%%%%%%%%%%

\section{Casimir-Polder shifts near an anisotropic slab}
In this section we are going to derive the energy shift for an atom close to a slab of material that is an anisotropic conductor or dielectric. We will tailor our results to the scenario of an atom interacting with a multilayer of graphene. To do so we take the electromagnetic response of the slab to be described by
\begin{eqnarray}
\epsilon_\parallel(\w)=1+i\frac{\sigma_\parallel(\w)}{\w},\;\;\;
\epsilon_\perp(\w)=n_\perp^2,
\label{eqn:Conductivity3}
\end{eqnarray}
that is, we assume that the slab behaves as a lossy conductor along its surfaces and as a non-dispersive and non-absorptive dielectric in the direction normal to its surfaces. Such a system loosely mimics a finitely thick wall of stacked graphene sheets where electrons are free to move along the sheets but their motion across the sheets is severely restricted.

\subsection{Ground state shifts}
The shift of the atomic ground state due to the presence of an anisotropic dielectric/conducting slab may be written down instantly by replacing the reflection coefficients in Eq.~(\ref{eqn:HalfSpaceShiftFinal}) with those appropriate for a slab, given in Eq.~(\ref{eqn:ReflectionCoeffSlab}). We get
\begin{eqnarray}
\Delta E_g=-\frac{1}{8\pi^2\epsilon_0}\sum_m\int_0^\infty\rd kk\int_{0}^\infty\rd\w\frac{\w_{mg}}{\w^2+\w_{mg}^2}\frac{e^{-2\sqrt{k^2+\w^2}\mathcal{Z}}}{\sqrt{k^2+\w^2}}\nonumber\\
\times\left\{\left[(k^2+\w^2)\bar{R}^\TM-\w^2\bar{R}^\TE\right]|\mu_{mg}^\parallel |^2+2k^2\bar{R}^\TM|\mu_{mg}^\perp|^2\right\},\nonumber\\\label{eqn:SlabShiftFinal}
\end{eqnarray}
where $\bar{R}$, apart from a phase factor, is the reflection coefficient for a slab; expressed in terms of the appropriate variables it reads
\begin{equation}
\bar{R}^\lambda=\bar{r}^\lambda\dfrac{1-e^{-2\bar{k}_{zd}^\lambda L}}{1-\bar{r}^2_\lambda e^{-2\bar{k}_{zd}^\lambda L}}.
\end{equation}
Here $\bar{r}_\lambda$ are the same as in Eqs.~(\ref{eqn:BarredFresnelTE}, \ref{eqn:BarredFresnelTM}) and the perpendicular components of the wave vectors in the dielectric medium $\bar{k}_{zd}^\lambda$ are
\begin{eqnarray}
\bar{k}_{zd}^\TE &=& \sqrt{\epsilon_\parallel(i\w)\w^2+k^2},\nonumber\\
\bar{k}_{zd}^\TM &=& \sqrt{\frac{\epsilon_\parallel(i\w)}{\epsilon_{\perp}(i\w)}}\sqrt{\epsilon_\perp(i\w)\w^2+k^2}.\nonumber
\end{eqnarray}
Later we are going to evaluate the energy shift numerically, whence we wish to rewrite Eq.~(\ref{eqn:SlabShiftFinal}) in a form which is more suitable for numerical analysis. To this end we go to polar coordinates, $\w=x\w_{mg}\cos\phi,\;k=x\w_{mg }\sin\phi$, and then set $\cos\phi=y$. After a short calculation we obtain
\begin{equation}
\Delta E_g=-\frac{1}{8\pi^2\epsilon_0\mathcal{Z}^4}\sum_m\frac{1}{\w_{mg}}\left[F^\parallel |\mu_{mg}^\parallel |^2+F^\perp|\mu_{mg}^\perp |^2\right]\label{eqn:SlabFinal2Anisotropic}
\end{equation}
with $|\mu_{mg}^\parallel |^2=|\mu_{mg}^x |^2+|\mu_{mg}^y |^2$ and
\begin{eqnarray}
F^\parallel &=&\int_0^\infty\rd xx^3\int_{0}^1\rd y \frac{(\mathcal{Z}\w_{mg})^4}{1+x^2y^2}
\left(\widetilde{R}^\TM-y^2\widetilde{R}^\TE\right)e^{-2\mathcal{Z}\w_{mg}x},\nonumber\\\label{eqn:FParaSlab}\\
F^\perp &=&\int_0^\infty\rd xx^3\int_{0}^1\rd y \frac{(\mathcal{Z}\w_{mg})^4}{1+x^2y^2}\left(1-y^2\right)2\widetilde{R}^\TM e^{-2\mathcal{Z}\w_{mg}x}.\nonumber\\\label{eqn:FPerpSlab}
\end{eqnarray}
Here the reflection coefficients are now given by
\begin{equation}
\widetilde{R}^\lambda=\widetilde{r}^\lambda\dfrac{1-e^{-2\widetilde{k}_{zd}^\lambda L}}{1-\widetilde{r}^2_\lambda e^{-2\widetilde{k}_{zd}^\lambda L}}.
\end{equation}
with $\widetilde{r}_\lambda$ being the single-interface coefficients listed in Eq.~(\ref{eqn:TildedFresnelAnisotropc}). The wave vectors in the dielectric medium $\widetilde{k}_{zd}^\lambda$ are now
\begin{eqnarray}
\widetilde{k}_{zd}^\TE &=& x\w_{mg} \sqrt{\left[\epsilon_\parallel(ixy\w_{mg})-1\right]y^2+1},\nonumber\\
\widetilde{k}_{zd}^\TM &=& x\w_{mg} \sqrt{\frac{\epsilon_\parallel(ixy\w_{mg})}{\epsilon_{\perp}(ixy\w_{mg})}}\sqrt{\left[\epsilon_\perp(ixy\w_{mg})-1\right]y^2+1}.\nonumber
\end{eqnarray}
Within the system of units we use in this paper ($\hbar=1=c$) the combinations $\mathcal{Z}\w_{mg}$ and $L\w_{mg}$ are dimensionless quantities, so that the functions $F_{\parallel,\perp}$ are also dimensionless.

\subsubsection{Nonretarded limit}\label{sec:Level5a}
Restoring the missing factors of the speed of light $c$ in Eq.~(\ref{eqn:SlabShiftFinal}), just as we did in Sec. \ref{sec:Lev4nonret}, and taking the limit $c\rightarrow\infty$, we find that when the atom is close to the interface the energy-shift of the ground state may be approximated by the formula
\begin{eqnarray}
\Delta E_g^{\rm nonret}&=&-\frac{1}{8\pi^2\epsilon_0}\sum_m\int_0^\infty\rd kk^2e^{-2k\mathcal{Z}}\int_0^\infty\rd \w\;\mathcal{R}(k,\w)\nonumber\\
&\times & \frac{\w_{mg}}{\w^2+\w_{mg^2}}\left(\left|\mu_\parallel\right|^2+\left|\mu_\perp\right|^2\right)\label{eqn:SlabNonRet}
\end{eqnarray}
where we have defined 
\begin{eqnarray}
\mathcal{R}(k,\w)=\dfrac{\sqrt{\epsilon_\parallel\epsilon_\perp}-1}{\sqrt{\epsilon_\parallel\epsilon_\perp}+1}\dfrac{1-e^{-2\sqrt{\epsilon_\parallel/\epsilon_\perp}kL}}{1-\left(\dfrac{\sqrt{\epsilon_\parallel\epsilon_\perp}-1}{\sqrt{\epsilon_\parallel\epsilon_\perp}+1}\right)^2e^{-2\sqrt{\epsilon_\parallel/\epsilon_\perp}kL}}.\nonumber
\end{eqnarray}
Here  $\epsilon_{\parallel,\perp}=\epsilon_{\parallel,\perp}(i\w)$ are evaluated at imaginary frequency and are completely general. The non-retarded limit $\mathcal{Z}\w_{mg}\rightarrow 0$ commutes with the mathematical limit pertaining to dielectric-conductor transition, $\mathcal{\w_{{\rm T}\parallel,\perp}}\rightarrow 0$, as well as with the limit of no damping, $ \gamma_{\parallel,\perp}\rightarrow0$. In particular, as long as we remain in the non-retarded or van der Waals regime, we may apply formula (\ref{eqn:SlabNonRet}) just as well with a dielectric response as in Eq.~(\ref{eqn:Conductivity}) as with that of Eq.~(\ref{eqn:Conductivity3}).

\subsubsection{Retarded limit}\label{sec:Level5b}
In order to find an asymptotic expansion of the energy shift in the retarded regime we take the limit $Z\w_{mg}\rightarrow\infty$ in Eq.~(\ref{eqn:SlabFinal2Anisotropic}). However, this limit does not commute with the dielectric-to-conductor limit $\mathcal{\w_{{\rm T}\parallel,\perp}}\rightarrow 0$ nor with the no-damping limit $ \gamma_{\parallel,\perp}\rightarrow0$, so that we must specify the electromagnetic response of the slab from the outset of the calculation. Let us restrict ourselves to the case when the slab has the optical properties given by Eq. (\ref{eqn:Conductivity3}). For large $\mathcal{Z}\w_{mg}$ the exponential factor in Eqs.~(\ref{eqn:FParaSlab}) and (\ref{eqn:FPerpSlab}) strongly damps the integrand so that the major contribution to the $x$ integral comes from the vicinity of the point $x=0$. Taylor-expanding the rest of the integrand around this point and then carrying out the $x$ and $y$ integrations, we find that, to leading order, the functions $F^{\parallel,\perp}$ entering Eq.~(\ref{eqn:SlabFinal2Anisotropic}) behave as
\begin{eqnarray}
F^\parallel &\approx & \frac{1}{8} +3\bigg\{\frac{1}{2L^2\sigma^2_\parallel(0)} -\frac{1}{8L\sigma_\parallel(0)}\nonumber\\
&+&\frac{L\sigma_\parallel(0)}{16}\ln\left[1+\dfrac{2}{L\sigma_\parallel(0)}\right]\nonumber\\
&-&\frac{1}{L^3\sigma^3_\parallel(0)}\ln\left[1+\dfrac{L\sigma_\parallel(0)}{2}\right]\bigg\},\nonumber\\
F^\perp &\approx &\frac{3}{8}L\sigma_\parallel(0)\bigg\{\frac{L\sigma_\parallel(0)-1}{2}\nonumber\\
&+& \left[1-L^2\sigma^2_\parallel(0)\right]\ln\left[1+\dfrac{2}{L\sigma_\parallel(0)}\right]\bigg\}.\label{eqn:SlabRetAsym}
\end{eqnarray}
where $\sigma_\parallel(0)=\w^2_{{\rm P}\parallel}/2\gamma_\parallel$ is the conductivity of the slab along its surfaces evaluated at zero frequency. 
We note that, unlike in the case of an atom interacting with an anisotropic half-space, cf.~Eq.~(\ref{eqn:retardedHSAnisotropic3}), the leading term of the asymptotic expansion in the retarded regime does depend on the optical properties of the slab (although it would independent of them if we had considered a lossless conductor). This interesting comparison demonstrates that cavity QED calculations based on artificial constructs such as a half-space, i.e.~an infinitely deep conductor, can be potentially misleading. 

If, instead of a conductor, we consider a slab made of a dielectric material (i.e.~a material with a dielectric constant of the form Eq.~(\ref{eqn:Coincide}) with non-zero 
$\omega_{\rm T}$) then, using the same procedure as described at the beginning of this section, we find that the leading-order terms of the asymptotic expansions of $F^{\parallel,\perp}$ in the retarded regime read
\begin{eqnarray}
F^\parallel\approx\frac{L}{Z}\left[\frac{9}{20}\epsilon_\parallel(0)-\frac{1}{4\epsilon_\perp(0)}-\frac{1}{5}\right],\nonumber\\
F^\perp\approx\frac{L}{Z}\left[\frac{1}{2}\epsilon_\parallel(0)-\frac{2}{5\epsilon_\perp(0)}-\frac{1}{10}\right].
\end{eqnarray}
where $\epsilon_{\parallel,\perp}(0)=1+\w^2_{{\rm P}\parallel,\perp}/\w^2_{{\rm T}\parallel,\perp}$ is the dielectric constant evaluated at zero frequency. This is in agreement with the common knowledge dating back to the work of McLachlan \cite{McLachlan} that it is the static response of the material that matters in the retarded regime of the Casimir-Polder interaction. By contrast, the dielectric response of an anisotropic conductor in Eq.~(\ref{eqn:Conductivity3}), which was used to derive the asymptotic expansion in Eq.~(\ref{eqn:SlabRetAsym}), behaves in the static limit as
\begin{eqnarray}
\epsilon_\parallel(\w\rightarrow 0)\rightarrow \infty,\;\;\;\epsilon_\perp(\w\rightarrow 0)=n_\perp^2.\label{eqn:StaticLimit}
\end{eqnarray}
On the basis of that one may be tempted to expect that in the retarded regime the leading-order behaviour of the energy shift near a conducting slab with an $\epsilon(\w)$ as in Eq.~(\ref{eqn:Conductivity3}) should be that of a perfect reflector, but Eq.~(\ref{eqn:SlabRetAsym}) has shown otherwise. The static dielectric response in Eq.~(\ref{eqn:StaticLimit}) facilitates perfect reflection for a half-space but not for a slab, as one can easily check by taking the zero-frequency limits of the appropriate reflection coefficients. For atoms interacting with conductors it is the static conductivity $\sigma_\parallel(0)$ that matters, and not $\epsilon_\parallel(0)$. Since $\sigma_\parallel(0)$ depends on the damping constant $\gamma_\parallel$ we conclude that damping plays an important role in the retarded Casimir-Polder interaction between an atom and a realistic conductor.

As a final remark we note that the leading-order term of the asymptotic series in the retarded regime is independent of $n_\perp$ and therefore our results and conclusions trivially extend to an isotropic conductor. Just as in the case of an atom interacting with a half-space, the response of the material in the direction normal to the slab's interfaces is to leading order irrelevant in the retarded regime, but it does very much matter in the non-retarded regime, as shown by Eq.~(\ref{eqn:SlabNonRet}).

\subsection{Numerical examples}
We now proceed to give a few numerical examples which illustrate the impact of the anisotropy of the material on the Casimir-Polder interaction between a neutral atom in its ground state and an anisotropic dielectric slab. For numerical purposes Eq.~(\ref{eqn:SlabFinal2Anisotropic}) is the most suitable. The numerical integration is straightforward as the integrals converge quickly because of the damping provided by the exponentials, and it can be carried out with standard computer algebra packages. This ceases to be the case only when the atom is very close to the surface but under such circumstances our model would not be valid anyway because the main effect of interaction would then be due to the overlap of the atomic wave function with that of the solid and thus of an entirely different nature to what has been investigated here.
%%%%%%%%%%%%%%%%%%%%%%%%%%%%%%%%%%%%%
\begin{figure}[t]
\includegraphics[width=8.5 cm, height=6 cm]{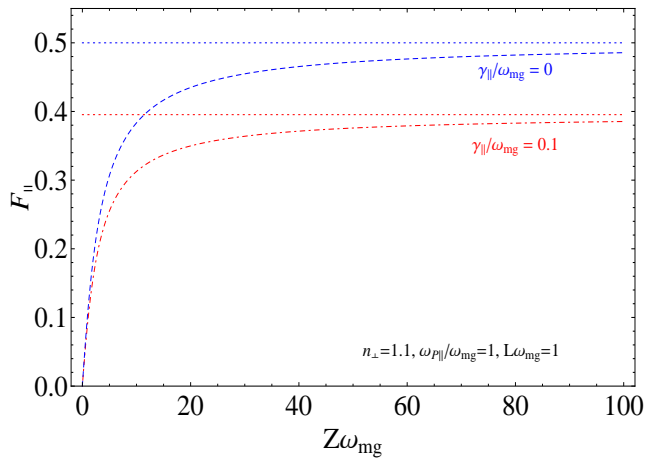}% Here is how to import EPS art
\caption{\label{fig:VerifyAsym} (Color online) Exact values of $F^\parallel$ of Eq.~(\ref{eqn:FParaSlab}) with $\epsilon(\w)$ as in Eq.~(\ref{eqn:Conductivity3}), with and without the losses in the material. In the case of no damping (blue, dashed) $F^\parallel$ approaches a constant value of $1/2$ for increasing $\mathcal{Z}\w_{mg}$, which is how a perfect reflector would behave in the retarded regime. When the material is absorbing (red, dot-dashed) $F^\parallel$ still approaches a constant value (i.e. the energy shift still vanishes as $1/\mathcal{Z}^4$) but the proportionality factor depends on the dimensionless combination $L\sigma_\parallel(0)$, cf. Eq. (\ref{eqn:SlabRetAsym}). }
\end{figure}
%%%%%%%%%%%%%%%%%%%%%%%%%%%%%%%%%%%%%%%%

To verify the asymptotic analysis of the energy shift of Eq.~(\ref{eqn:SlabFinal2Anisotropic}) in the retarded regime, Sec.~\ref{sec:Level5b}, we plot in Fig.~\ref{fig:VerifyAsym} the exact value of $F^\parallel$ with and without losses in the material. In the case of no damping (blue, dashed) we observe that as $\mathcal{Z}\w_{mg}$ increases $F^\parallel$ approaches a constant value of $1/2$, which indicates that, to leading order, the energy shift decays as $1/Z^4$ with the atom-surface separation, with a factor of proportionality that is the same as in the case of a perfectly reflecting surface. This confirms our remarks made in Sec.~\ref{sec:Level5b} regarding the lossless slab. When the material is absorbing (red, dot-dashed) the function $F^\parallel$ still approaches a constant value i.e.~the energy shift still vanishes as $1/\mathcal{Z}^4$, but the factor of proportionality (red, dotted) depends on the static conductivity of the slab along its surfaces and its thickness through the dimensionless combination $L\sigma_\parallel(0)$.

%%%%%%%%%%%%%%%%%%%%%%%%%%%%%%%%%%%%%
\begin{figure}[ht]
\includegraphics[width=8.5 cm, height=6 cm]{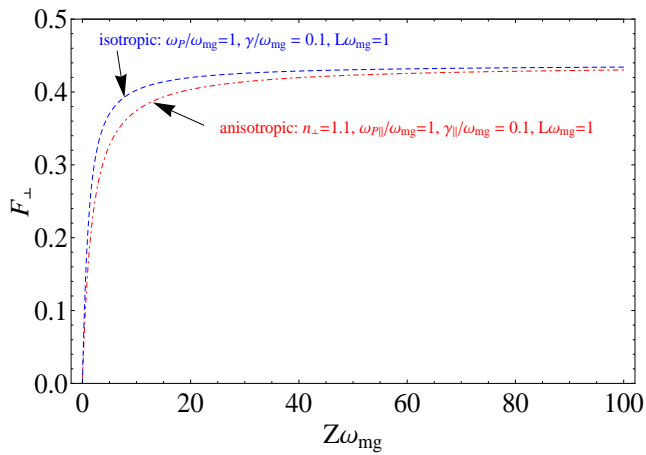}% Here is how to import EPS art
\caption{\label{fig:AnisotropySlabRet} (Color online) The function $F^\perp$ of Eq.~(\ref{eqn:FPerpSlab}) for an isotropic conductor (blue, dashed) and for an $\epsilon(\w)$ as in Eq.~(\ref{eqn:Conductivity3}), where the conductivity of the slab in the direction normal to the surface is suppressed and the slab's optical response in this direction is that of a non-dispersive dielectric (red, dot-dashed). As $\mathcal{Z}\w_{mg}$ increases the difference between the energy shifts in the two cases vanishes, which confirms our conclusion that the anisotropy of the conductor does not matter in the retarded regime.}
\end{figure}
%%%%%%%%%%%%%%%%%%%%%%%%%%%%%%%%%%%%%%%%

\begin{figure}[ht]
\includegraphics[width=8.5 cm, height=6 cm]{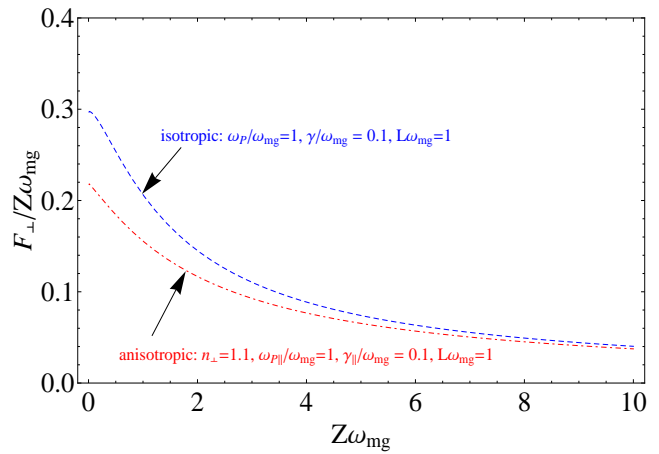}% Here is how to import EPS art
\caption{\label{fig:AnisotropySlabNonRet} (Color online) The function $F^\perp/\mathcal{Z}\w_{mg}$ of Eq.~(\ref{eqn:FPerpSlab}) for an isotropic conductor (blue, dashed) and for an anisotropic conductor with an $\epsilon(\w)$ as in Eq.~(\ref{eqn:Conductivity3}) (red, dot-dashed). For small distances $\mathcal{Z}\w_{mg}$ the anisotropy of the material has a considerable impact on the Casimir-Polder interaction. In this example, the Casimir-Polder force at $\mathcal{Z}\w_{mg}\approx 1$ is reduced by about $25\%$ due to
the suppression of the conductivity in the direction normal to the slab's surface.}
\end{figure}

Next we would like to illustrate the impact of the material's anisotropy on the Casimir-Polder interaction. For this purpose we plot in Fig.~\ref{fig:AnisotropySlabRet} the function $F^\perp$ of Eq.~(\ref{eqn:FPerpSlab}) for two different cases. In the first case we take the material to be an isotropic conductor (blue, dashed), and in the second case we assume that the conductivity of the slab in the direction normal to the surface is suppressed and that the slab's optical response in this direction is that of a non-dispersive dielectric (red, dot-dashed). Fig. \ref{fig:AnisotropySlabRet} illustrates that in the retarded regime the anisotropy of the material becomes irrelevant. However, Fig.~\ref{fig:AnisotropySlabRet} is not convenient for looking at the energy shift in the non-retarded regime, where it is much more convenient to plot $F^\perp/\mathcal{Z}\w_{mg}$ because this combination approaches a constant value for $\mathcal{Z}\w_{mg}\rightarrow0$. We do just that in Fig.~\ref{fig:AnisotropySlabNonRet}, which shows that the anisotropy of the conductor strongly affects the Casimir-Polder interaction in the non-retarded regime. 

%%%%%%%%%%%%%%%%%%%%%%%%%%%%%%%%%%%%%
%%%%%%%%%%%%%%%%%%%%%%%%%%%%%%%%%%%%%%%%

\section{Summary and conclusions}
The quantum propagator of the displacement field describes the propagation of photons in the multipolar formulation of nonrelativistic QED. Modelling an anisotropic dielectric as a set of quantized harmonic oscillators coupled to a heat bath we have successfully derived the quantum photon propagator near an anisotropic slab, which may be taken to be either a conductor or a dielectric, or even one type in one direction and another type perpendicular to it. We have used the photon propagator to work out the energy-level shifts for a variety of scenarios, but perhaps most importantly for an atom close to a slab that in the direction parallel to its surface behaves as lossy conductor and in the direction normal to it as a nondispersive and non-absorptive dielectric. We have found that the anisotropy of the material matters most in the non-retarded regime and that it affects the Casimir-Polder interaction considerably when the atom is close to the surface. In the opposite case, when the atom is far away from the surface, the significance of the material's anisotropy depends on the optical response of the material one deals with. For atoms interacting with anisotropic conductors the anisotropy does not matter to leading order, but it does matter for anisotropic dielectrics. This is in contrast to the role of damping in the retarded regime: it does not matter for dielectrics but it is important for conductors. Our results for the energy shift may be used to estimate the Casimir-Polder force acting on quantum objects trapped close to multilayers of graphene or graphite. Our results are particularly important for the case of cold molecules whose dispersive interactions with surfaces often fall within the non-retarded regime where the anisotropy of the material strongly affects the Casimir-Polder force.

\begin{acknowledgments}
It is a pleasure to thank Mark Fromhold for discussions. 
We would like to acknowledge financial support from the UK Engineering and
Physical Sciences Research Council.
\end{acknowledgments}

\appendix
\section{Excited-state energy shifts near an anisotropic half-space\label{sec:App1}}
The additional energy shift of an excited state $|i\rangle$ (and the spontaneous decay rate as its imaginary part) can be obtained by using formula (\ref{eqn:ShiftFinal1}) with the propagator of Eq.~(\ref{eqn:ReflectedPropagatorAnisotropic}). The general formula for the additional energy-level shift is given by
\begin{eqnarray}
&&\Delta \mathcal{E}_i^{\star}=\frac{i}{8\pi\epsilon_0}\sum_{m<i}|\w_{mi}|^3\int_1^{i\infty}\rd \kappa\; e^{2i|\w_{mi}|\mathcal{Z}\kappa} \label{eqn:ResidueContributions1Anisotropic}\\
&&\hspace*{5mm}\times\left\{ \left(\bar{r}^\TE_{mi}-\kappa^2\bar{r}^\TM_{mi}\right)|\mu_{mi}^\parallel|^2+2\left(1-\kappa^2\right)\bar{r}^\TM_{mi}|\mu_{mi}^\perp|^2\right\}.\hspace{.7 cm}
\nonumber
\end{eqnarray}
where $|\mu_{mi}^\parallel |^2=|\mu_{mi}^x |^2+|\mu_{mi}^y |^2$, and $|\w_{mi}|\equiv |\w_m-\w_i|$ is the modulus of the transition frequency between the states $|i\rangle$ and $|m\rangle$. The dimensionless variable $\kappa$ depends on the original $q_\parallel$ through $\kappa=\sqrt{\w_{mi}^2-q_\parallel^2}/|\w_{mi}|$ and its contour of integration runs from $\kappa=1$ along the real axis to $\kappa=0$ and then up along the imaginary axis to $\kappa=i\infty$. The reflection coefficients expressed as functions of $\kappa$ are
\begin{eqnarray}
\bar{r}^\TE_{mi}(\kappa)&=&\frac{\kappa-\sqrt{\epsilon_\parallel-1+\kappa^2}}{\kappa+\sqrt{\epsilon_\parallel-1+\kappa^2}},\nonumber\\
\bar{r}^\TM_{mi}(\kappa)&=&\frac{\sqrt{\epsilon_\parallel\epsilon_\perp}\kappa-\sqrt{\epsilon_\perp-1+\kappa^2}}{\sqrt{\epsilon_\parallel\epsilon_\perp}\kappa+\sqrt{\epsilon_\perp-1+\kappa^2}}.\label{eqn:ChangedFresnels}
\end{eqnarray}
Here the dielectric permittivities are evaluated at the real frequency $|\w_{mi}|$, i.e.~$\epsilon_{\sigma}=\epsilon_\sigma(|\w_{mi}|)$. The asymptotic analysis of Eq.~(\ref{eqn:ResidueContributions1Anisotropic}) can be carried out by exactly the same methods as those described in Sec. VI B of Ref.~\cite{Dispersive1}. Therefore we merely quote the results.

\paragraph{Nonretarded limit:}
In the limit of instantaneous interaction, $2|\w_{mi}|\mathcal{Z}\rightarrow 0$, Eq.~(\ref{eqn:ResidueContributions1Anisotropic}) becomes
\begin{eqnarray}
\Delta \mathcal{E}_i^{\star,{\rm nonret}}&\approx &-\frac{1}{32\pi\epsilon_0\mathcal{Z}^3}\sum_{m<i}\frac{\sqrt{\epsilon_\parallel(|\w_{mi}|)\epsilon_\perp(|\w_{mi}|)}-1}{\sqrt{\epsilon_\parallel(|\w_{mi}|)\epsilon_\perp(|\w_{mi}|)}+1}
\nonumber\\
&\times & \left(|\mu_{mi}^\parallel|^2+2|\mu_{mi}^\perp|^2\right).\;\;\;\label{eqn:ExcitedNonRet}
\end{eqnarray}
whose real part gives the residue contributions to the energy shift of the excited state $|i\rangle$ to leading order as
\begin{eqnarray}
\Delta E_i^{\star, {\rm nonret}}&\approx &-\frac{1}{32\pi\epsilon_0\mathcal{Z}^3}\sum_{m<i}\frac{|\epsilon_\parallel(|\w_{mi}|)\epsilon_\perp(|\w_{mi}|)|-1}{|\sqrt{\epsilon_\parallel(|\w_{mi}|)\epsilon_\perp(|\w_{mi}|)}+1|^2}\nonumber\\
&\;&\times\left(|\mu_{mi}^\parallel|^2+2|\mu_{mi}^\perp|^2\right).\;\;\;\label{eqn:ReExcitedNonRet}
\end{eqnarray}
Thus, in the nonretarded regime the residue contributions depend on distance as $\mathcal{Z}^{-3}$ and therefore are of the same order as the ground-state shifts.

\paragraph{Retarded limit:}
In the retarded limit Eq.~(\ref{eqn:ResidueContributions1Anisotropic}) becomes
\begin{eqnarray}
\Delta \mathcal{E}_i^{\star \rm,ret}&\approx &\frac{1}{4\pi\epsilon_0} \sum_{m<i}|\w_{mi}|^3\frac{n_\|(|\w_{mi}|)-1}{n_\|(|\w_{mi}|)+1}\;e^{2i|\w_{mi}|\mathcal{Z}}
\nonumber\\
&\times &\left[\frac{|\mu_{mi}^\parallel|^2}{2|\w_{mi}|\mathcal{Z}} +2i\frac{|\mu_{mi}^\perp|^2}{(2|\w_{mi}|\mathcal{Z})^2}\right],\;\;\;\label{eqn:ExcitedRet}
\end{eqnarray}
with the parallel refractive index $n_\|(|\w_{mi}|)=\sqrt{\epsilon_\parallel(|\w_{mi}|)}$. Thus, the energy-level shifts, given by the real part of the above expression, read
\begin{eqnarray}
\Delta E_i^{\rm \star,ret}\approx\frac{1}{4\pi\epsilon_0}\sum_{m<i}\frac{|\w_{mi}|^3}{|n_\|(\w_{mi})+1|^2}\hspace{3 cm}
\nonumber\\
\times\bigg\{\bigg[(|n_\|(|\w_{mi}|)|^2-1)\cos(2|\w_{mi}|\mathcal{Z})\hspace{2.8 cm}
\nonumber\\
-2{\rm Im}[n_\|(|\w_{mi}|)]\sin(2|\w_{mi}|\mathcal{Z})\bigg]\frac{|\mu_{mi}^\parallel|^2}{2|\w_{mi}|\mathcal{Z}}\hspace{1.5 cm}
\nonumber\\
-2\bigg[(|n_\|(|\w_{mi}|)|^2-1)\sin(2|\w_{mi}|\mathcal{Z})\hspace{2.8 cm}
\nonumber\\
+2{\rm Im}[n_\|(|\w_{mi}|)]\cos(2|\w_{mi}|\mathcal{Z})\bigg]\frac{|\mu_{mi}^\perp|^2}{(2|\w_{mi}|\mathcal{Z})^2}\bigg\}.\hspace{.7 cm}\label{eqn:ExitedFinalRetAnisotropic}
\end{eqnarray}

\section{Spontaneous decay rates near an anisotropic half-space\label{sec:App2}}
In the non-retarded limit the spontaneous decay rates are given by the imaginary part of Eq.~(\ref{eqn:ExcitedNonRet})
\begin{eqnarray}
\Delta\Gamma^{\rm nonret}_i &=& \frac{1}{8\pi\epsilon_0\mathcal{Z}^3}\sum_{m<i}\frac{{\rm Im}[\sqrt{\epsilon_\parallel(|\w_{mi}|)\epsilon_{\perp}(|\w_{mi}|)}]}{|\sqrt{\epsilon_\parallel(|\w_{mi}|)\epsilon_{\perp}(|\w_{mi}|)}+1|^2}\nonumber\\
&\;& \times \left(|\mu_{mi}^\parallel|^2+2|\mu_{mi}^\perp|^2\right),\label{eqn:HSDecayNonRet}
\end{eqnarray}
whereas in the retarded limit by the imaginary part of Eq.~(\ref{eqn:ExcitedRet}):
\begin{eqnarray}
\Delta \Gamma^{\rm ret}_i=-\frac{1}{2\pi\epsilon_0}\sum_{m<i}\frac{|\w_{mi}|^3}{|n_\|(\w_{mi})+1|^2}\hspace{2.8 cm}\nonumber\\
\times\bigg\{\bigg[(|n_\|(|\w_{mi}|)|^2-1)\sin(2|\w_{mi}|\mathcal{Z})\hspace{2.5 cm}\nonumber\\
+2{\rm Im}[n_\|(|\w_{mi}|)]\cos(2|\w_{mi}|\mathcal{Z})\bigg]\frac{|\mu_{mi}^\parallel|^2}{2|\w_{mi}|\mathcal{Z}}\hspace{0.8 cm}\nonumber\\
+2\bigg[(|n_\|(|\w_{mi}|)|^2-1)\cos(2|\w_{mi}|\mathcal{Z})\hspace{2.5 cm}\nonumber\\
-2{\rm Im}[n_\|(|\w_{mi}|)]\sin(2|\w_{mi}|\mathcal{Z})\bigg]\frac{|\mu_{mi}^\perp|^2}{(2|\w_{mi}|\mathcal{Z})^2}\bigg\},\hspace{0.6 cm}\label{eqn:HSDecayRet}
\end{eqnarray}
with the parallel refractive index $n_\|(|\w_{mi}|)=\sqrt{\epsilon_\parallel(|\w_{mi}|)}$. The behaviour of the excited-state shifts and spontaneous decay rates in the retarded regime is very much different from the energy shift of the ground state in the far zone because they oscillate. This difference can be explained by the fact that the ground-state shifts are caused by virtual photons whereas an excited atom may emit real photons. Therefore, the shift of an excited level bears a close analogy to a classical dipole oscillating with the frequency $|\w_{mi}|$ near an interface. The analogy persists for an anisotropic half-space. Note that the results (\ref{eqn:ExitedFinalRetAnisotropic}) and (\ref{eqn:HSDecayRet}) do not, to leading order, depend on $\epsilon_\perp(\w)$. This is plausible if one notes that the electric field of a radiating dipole that is sufficiently far away from a surface is to a good approximation parallel to the surface. Thus it is insensitive to the response of the material in the direction perpendicular to the surface.


\begin{thebibliography}{99}
\bibitem{fn1}{The transition from a dielectric to a conductor is achieved by setting the potential of the restoring force in Eq.~(\ref{eqn:PolarizationHam}) to zero, i.e.~$\w_{\rm T}\rightarrow 0$. The conductor is then envisaged as a free-electron gas with an overall-neutralizing background like in the hydrodynamic model of metals.}
\bibitem{fn2}{Most of the integrals in Eq.~(\ref{eqn:FFunctions2}) can in principle be computed analytically leading to expressions involving Bessel and related functions, but the result is not nearly as useful as the integral representation (\ref{eqn:FFunctions2}).}
\end{thebibliography}
\end{document}